\documentclass{aa}  
\usepackage{graphicx}
\usepackage{txfonts}
\usepackage{xspace}
\usepackage{color}
\usepackage{hyperref}

\newcommand{\mysp}{}\def\mysp/{}

\newcommand{\forbm}[3]{[\ion{#1\mysp/}{#2}]\ #3$\mu$m\xspace}

\newcommand{\dforbm}[4]{[\ion{#1\mysp/}{#2}]\ #3,#4$\mu$m\xspace}

\newcommand{\mm}{$\mu$m\xspace}

\newcommand\T{\rule{0pt}{2.6ex}}       % Top strut
\newcommand\B{\rule[-1.2ex]{0pt}{0pt}} % Bottom strut

\begin{document}

   \title{Modeling dust emission in PN IC~418 \thanks{Based on observations with ISO, an ESA project with instruments funded by ESA Member States (especially the PI countries: France, Germany, The Netherlands and the UK) and with the participation of ISAS and NASA.}}

   \subtitle{}

   \author{V. G\'omez-Llanos\inst{1}
          \and
          C. Morisset\inst{1} 
          \and 
          R. Szczerba\inst{2} 
          \and
          D. A. Garc\'{\i}a-Hern\'andez\inst{3, 4}
          \and 
          P. Garc\'{\i}a-Lario\inst{5}
          }

   \offprints{V. Gomez-Llanos and C. Morisset}

   \institute{Instituto de Astronom\'{\i}a,
     Universidad Nacional Aut\'onoma de M\'exico, 
     Apdo. postal 70--264; 
     Ciudad Universitaria; M\'exico CDMX 04510; M\'exico. \\ 
     \email{vgomez.astro@gmail.com}, ORCID=0000-0002-1825-8267, \email{Chris.Morisset@gmail.com}, ORCID=0000-0001-5801-6724.
     \and 
     Nicolaus Copernicus Astronomical Center, PAS, ul. Rabia\'nska 8, 87-100 Toru\'n, Poland
     \and
     Instituto de Astrof\'isica de Canarias, V\'ia L\'actea s/n, E-38205 La Laguna, Tenerife, Spain
     \and
     Departamento de Astrof\'isica, Universidad de La Laguna (ULL), E-38206 La Laguna, Tenerife, Spain
     \and
    Herschel Science Centre, ESAC/ESA, PO Box 78, E-28691 Villanueva de la Cañada, Madrid, Spain
      }

   \date{Received; accepted}
 
  \abstract{We investigated the infrared (IR) dust emission from PN IC\,418, using a detailed model controlled by a previous determination of the stellar properties and the characteristics of the photoionized nebula, keeping as free parameters the dust types, amounts and distributions relative to the distance of the central star. 
The model includes the ionized region and the neutral region beyond the recombination front (Photodissociation region, or PDR), where the [OI] and [CII] IR lines are formed. 
We succeeded in reproducing the observed infrared emission from 2 to 200~\mm. The global energy budget is fitted by summing up contributions from big grains of amorphous carbon located in the neutral region and small graphite grains located in the ionized region (closer to the central star).
   Two emission features seen at 11.5 and 30~\mm are also reproduced by assuming them to be due to silicon carbide (SiC) and magnesium and iron sulfides (Mg$_x$Fe$_{1-x}$S), respectively. For this, we needed to consider ellipsoidal shapes for the grains to reproduce the wavelength distribution of the features.
   Some elements are depleted in the gaseous phase: Mg, Si, and S have sub-solar abundances (-0.5 dex below solar by mass), while the abundance of C+N+O+Ne by mass is close to solar. Adding the abundances of the elements present in the dusty and gaseous forms leads to values closer to but not higher than solar, confirming that the identification of the feature carriers
is plausible. Iron is strongly depleted (3 dex below solar) and the small amount present in dust in our model is far from being enough to recover the solar value.
A remaining feature is found as a residue of the fitting process, between 12 and 25~\mm, for which we do not have identification.
   }


   \maketitle
%
%________________________________________________________________

\section{Introduction}\label{sec:introduction}
Planetary nebulae (PNe), as the ultimate stage of intermediate mass stars, are an essential provider in the chemical enrichment of the Universe. They also generate, while in the asymptotic giant branch (AGB) phase previously to the PN, most of the dust present in the local Universe \citep[e.g.,][]{2006Ferrarotti_aap447}. This dust occults some elements from the observer, diminishing their
emission in the gas phase. When studying the chemical evolution by obtaining the total abundances in the interstellar medium (ISM) it is necessary to account for both gas and dust abundances. We note that obtaining the dust abundances is not an easy task.

The relatively simple morphology and the great number of observations on the PN \object{IC~418}, especially in the infrared (IR), makes it a good object to study the dust emission. 
In our paper we concentrate on the dust component of PN IC~418, which together with the very detailed analysis of atomic gas and stellar component in this nebula by 
\citet[][]{2009Morisset_aap507} (hereafter MG09) provide a 
quite complete model of this object. However, we do not model the molecular component, which to our best knowledge, is not observed in IC~418.

An analysis of near infrared (NIR) maps of IC~418 has been made by \citet{1986Phillips_apss122}, finding that this nebula will need at least two types of grains, one that will emit at longer wavelengths (with a larger grain radius) and another with a small radius, that causes the excess at the NIR. Two very interesting features are found in the infrared emission of IC~418. The first, found by \citet{1979Willner_apj234}, is an emission excess from 10.5 to 13~\mm, which they suggest is due to silicon carbide (SiC) grains, following \citet{1974Treffers_apj188}. The second is an emission feature at $\lambda \sim $ 30~\mm, first partially (in the 16 - 30~\mm range) found by \citet{1981Forrest_apj248} in a spectroscopic study of carbon stars and PNe. It is fully reported by \citet{1985NASCP2353..233M,1986NASCP2403A..18M} and usually associated to magnesium sulfide (MgS) grains, following \citet{1985Goebel_apjl290}. 

Modeling dust emission, in particular from \object{IC~418}, has been done by \citet{1990Hoare_mnra244}. A very detailed photoionization and stellar atmosphere model has been presented by MG09, that reproduces the main characteristics of the nebular and stellar emissions. More recently, \citet{2017Dopita_mnra470} present a revised model of the nebula, taking into account the presence of shocks at the inner and outer edges of the ionized region. 
In this paper we have carried out a new study of the IR dust emission for this object, mainly based on ISO observations and using as a starting point the MG09 model. 

Obtaining a model that fits the dust emission, in the global as well as in the detail, allow us to compute the amount of elements embedded in dust and compare them to the amount present in the gaseous phase. Given that \object{IC~418} is marginally over-solar for C+N+O+Ne (0.1 dex above the solar value, by mass, see MG09), we expect to obtain total abundances (gaseous phase + dust) for Mg, Si, S, and Fe comparable to the solar values.

The structure of this paper is as follows: in Section 2 we describe the observations of PN \object{IC~418} used to constrain the models. In Section 3 we present the model needed to reproduce the continuum IR dust emission. In Section 4 the detailed modeling of two special spectral features is shown. In Section 5 we present the best fit model. The discussion and conclusions are given in Sections 6 and 7, respectively.

\section{Observations}\label{sec:observations}

\object{The source IC~418} is a very well observed young low-ionization PN of size 10\arcsec x 12\arcsec. In this work we use observations from the literature and from public access archive.
The ISO observations were reduced using ISAP and LIA software with OLP version 10.1, where leakage from the 13 \mm region is corrected \citep[see][]{2002Hony_aap390}.
The spectrum up to 27.5 \mm was normalized to IRAS photometry at 12 and 25 \mm, and band 4 was scaled to smoothly join LWS spectrum which was normalized to the 100 \mm IRAS flux. Then band 3E between 27.5 and 29 \mm was scaled to join these two parts of the spectrum. 
We notice that the ISO smallest aperture is 14\arcsec x 20\arcsec, thus the whole PN is observed and no aperture effect is expected.
Spitzer data are only covering the 9.9-19.6 \mm range, and are in a relative good agreement with the ISO data (The Spitzer Infra-Red Spectrograph  Short-High spectrum was taken through an aperture size of 4.7\arcsec x 11.3\arcsec, then only part of the nebula is observed). In the following we only used the ISO data.

\citet{1981Forrest_apj248} give observed points from KAO, showing a double peak between 25 and 30\mm, which is also seen in the ISO data. However, \cite{2002Hony_aap390}, argued that such appearance of the 30 \mm\ feature is a result of the instrumental effect from the ISO spectrograph. It is not clear what is the reason for the double peak structure of the 30\,${\mu}$m feature from the KAO data. The fluxes reported are smaller than the ISO fluxes by a factor of $\sim$1.6. Aperture size effects are not suspected to be the cause of this discrepancy, as the KAO beam is 30'' wide.
The ISO spectroscopic data are in a very good agreement with the IRAS data at 12, 25, 60 and 100~\mm.
We also used far-IR and radio observations at 450, 800, 801, and 1100 \mm from \citet{1992Hoare_mnra258} and radio observations between 2700 and 353000 MHz from \citet{1970Dixon20, 1991Wright251, 2010Vollmer511, 2010Murphy402,2015Planck-Collaboration_aap573}.

%__________________________________________________________________

\section{The infrared continuum modeling}\label{sec:infra-red-contiuum}

Previous models of the IR emission of PN IC~418 have been made by \citet{1990Hoare_mnra244}, before ISO data were available. They used a model combining the emission of the H$^+$ fotoionized region and the emission from a neutral region.

In a previous work, MG09 compute a detailed 3D model of IC418, combining a stellar model using CMFGEN \citep{1998Hillier_apj496} and a photoionization model using Cloudy\_3D \citep{2006Morisset_234}, based on Cloudy code \citep{1998Ferland_pasp110}. The density distribution used in MG09 was not spherical, but rather axi-symmetrical, to reproduce the Hubble Space Telescope (HST) images and to carefully compute the line intensities as observed through different apertures depending on the instruments used. In the present work, we will not need this level of details because the observations we aim to fit have been obtained by ISO spectrometers, with aperture sizes comparable or greater than the size of IC~418. The departure from sphericity of IC418 is small and using a spherical model will decrease the CPU time without affecting our conclusions. We checked a posteriori that increasing the position of the two density shells by 10\% does not significantly change the result. A change of a maximum of 5\% in the dust emission is obtained, no change in the shape. The abundances of the elements in the gas are set as in the MG09 model, except for Si (see next section).
The models used in this study are obtained with Cloudy v. 17.00  \citep{2017Ferland_rmxaa53} and the pyCloudy library \citep{2014Morisset_}.

The photoionization models presented in this work are all obtained using the stellar atmosphere model presented in MG09 (with T$_{\rm eff}\sim$36.7~kK) but with a smaller luminosity of 5900 solar luminosity, as less dust is present in the ionized region of the present detailed model than assumed by MG09. We kept the distance to its MG09 value of 1.26~kpc. These luminosity and distance lead to a very good agreement between the observed and predicted values for the radio emission reported by \citet{1990Hoare_mnra244}. On the other side, the H$\beta$ flux is not so well reproduced by the model. The observed value is 2.7 10$^{-10}$ erg cm$^{-2}$ s$^{-1}$ \citep{1992Cahn_aaps94}.
Using a distance of 1.26~kpc, E$_{B-V} = 0.26$, and R$_V = 3.6$ \citep[MG09,][]{1999Fitzpatrick_pasp111}, a total emitted flux of 1.37~10$^{35}$ erg s$^{-1}$ is obtained, while the model predicts a value of 1.03 10$^{35}$ erg s$^{-1}$. We did not find any way to reconcile H$\beta$ and radio fluxes with the same model. In their recent model of IC418, \citet{2017Dopita_mnra470} are fitting H$\beta$, but they do not predict the IR nor the radio emission.
We actually prefer to fit the radio observations, less influenced by uncertainties on the reddening correction.

\subsection{Photoionization model extended into the PDR}\label{sec:phot-model-extend}
\begin{figure}
   \centering
 \includegraphics[width=9cm]{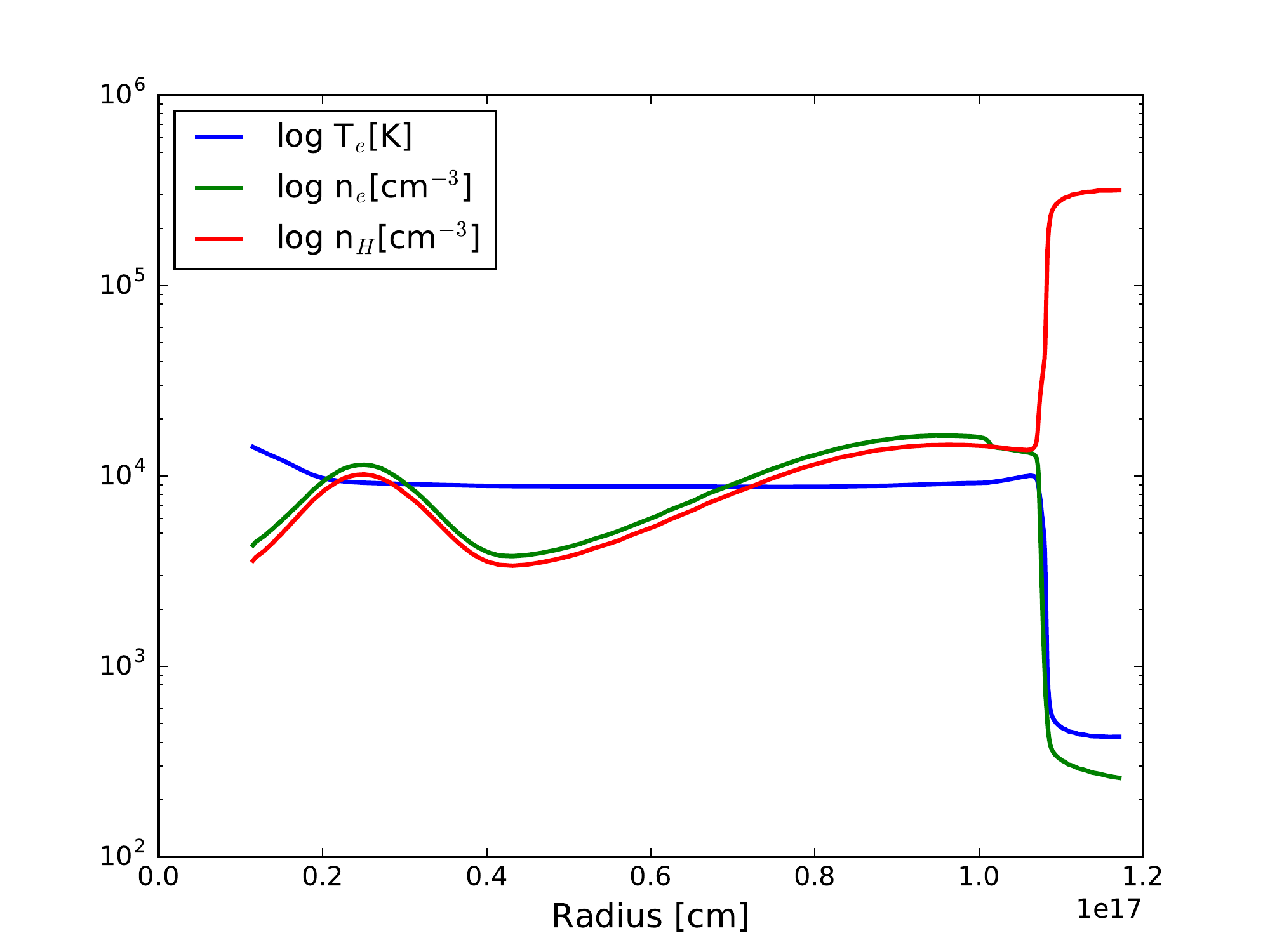}
   \caption{Electron temperature, electron density, and hydrogen density.}
              \label{fig:dens}%
\end{figure}

The H density distribution used in the models presented here is a mean between the short- and long-axis distribution used in MG09, namely a double shell, as shown in Fig.~\ref{fig:dens}.
In the model presented by MG09, the computation is stopped when H recombines, just behind the secondary peak. But the observation of strong fine structure IR lines issuing from the neutral region (namely \dforbm{O}{i}{63}{146} and \forbm{C}{ii}{158}) indicate that the nebula actually extends beyond the recombination front and that a PDR region is present. The PDR properties can be constraint using Fig.~6 of \citet{2001Liu_mnra323}, who determined a temperature of 220~K and a density of 3~10$^5$~H/cm$^3$ for IC418 (we check with the final model that none of these lines is optically thick and that the relation is valid). This PDR also contains dust taking part to the observed IR emission, and we needed to extend the photoionization models to include this region. This is achieved by allowing Cloudy to continue the computation after the recombination front. The PDR density is $\sim$20 times higher than the density of the ionized outer shell, while the electron temperature is $\sim$35 times lower. This lead us to adopt a very simple density law for the PDR of the form: n$_{\rm H}(r) = n_{\rm H}^0 \times \cdot T_e^0/T_e(r)$, for $r > R_{\rm recomb}$. We set $n_{\rm H}^0$ and $T_e^0$ to their values just before the recombination front, that is 13,000~H/cm$^3$ and 10,000~K respectively.

The thickness of the PDR is determined by stopping the computation when the intensity of one of the IR lines reaches its observed value: we used \forbm{O}{i}{63} absolute intensity of 5.2~10$^{-11}$ erg cm$^{-2}$ s$^{-1}$ \citep{2001Liu_mnra323,2004Pottasch_aap423} as stopping criterion. The model actually stopped at the closest value of 4.8~10$^{-11}$ erg cm$^{-2}$ s$^{-1}$. The value used to stop the computation (i.e., the size of the PDR) has a direct influence on the dust-to-gas ratio (D/G) we obtain by fitting the IR dust emission.
The PDR has a relatively small thickness ($\sim$1.5~10$^{15}$~cm, but see below), but contains a mass of gas of order of the ionized shell one (the H$^+$ mass is 0.047 M$_{\odot}$ while the H$^0$ mass is 0.039 M$_{\odot}$).
We don't think the PDR can be much bigger than this value, otherwise the intensities of these infrared lines would be quite higher than the observation, especially considering that the lines are not optically thick. We note also that \citet{1996Dayal_apj472} did not found evidence of CO emission from IC418, so that we ruled out the presence of a molecular halo, which could also have emitted some IR continuum.
The geometrical thickness of the PDR is very small compared to the size of the nebula (on the order of 1\%). We chose to add a filling factor to the PDR region so that this size increases. It does not change anything else regarding the line or dust emission, but it makes the plots easier to look at. We used a value for this PDR filling factor of 0.1.

We used the "iterate 3" option in the Cloudy model to ensure that the values for the IR lines are well computed.
The final model presented below has a mean electron temperature of $\sim$462~K and a mean hydrogen density of $\sim$3.0~10$^5$~cm$^3$ for the PDR, see Fig.~\ref{fig:dens} where we show the radial variations of the H and e densities, as well as the electron temperature. We clearly see the jump of all these quantities at the recombination front.

It turns out that \forbm{Si}{ii}{34.8}, the only available line of silicon, is mainly emitted by the PDR. To fit the observed value (relative to \forbm{O}{i}{63.17}), an abundance of only -5.55 (in log) is needed, while MG09 obtained -4.90 when taking only the ionized region into account.

The electron- (or even more relevant in the case of a PDR, proton-) temperature can be obtained from diagnostic diagrams following \citet{2001Liu_mnra323}. We note that since their paper, new atomic data have been published \citep[e.g.,][]{2007Abrahamsson_apj654} that would lead to an even lower temperature for the PN PDRs. The [CII]/[OI] line ratio diagnostic has to be normalized by the C/O. This latest value is known for the ionized part of the nebula, but the IR emission lines actually come from the H$^0$ region where C can be more depleted, changing the diagnostics.

Our model fails somewhat in reproducing the following diagnostics: the modeled (observed) intensity ratios for \forbm{O}{i}{145.5/63.17} and \forbm{C}{ii}{157.6}/\forbm{O}{i}{63.17} are: 0.028 (0.028) and 0.085 (0.112), respectively.
The PDR is then predicted systematically warmer than what determined from line ratios. But the dust temperature is totally decorrelated from the gas temperature. Not fitting the gaseous temperature does not jeopardize the dust emission model.

In a recent work, \citet{2017Dopita_mnra470} present new observations and a new model of the IC~418 nebula, based on the photoionization+shock modeling tool Mappings V \citep{2005Dopita_apj619,2017Sutherland_apjs229}. They found a global agreement with the model described by MG09. They added two shocked regions: one in the inner face of the H$^+$ region, due to the stellar wind ram pressure, and a second one in the outer part of the same H$^+$ region, where it collides with the preexisting AGB stellar wind. Contrary to the model presented in our work, their model is not able to take into account the PDR, which according to our model contains five times more dust than the ionized region (see Table\ref{tab:results}).
We were not able to include the shocked regions in our model, as Cloudy does not manage them. Nevertheless, we prefer including the PDR than the shocked region.

\subsection{Usual dust emission is too hot: need for cooler grains}\label{sec:usual-<-emission}

To reproduce the global shape of the IR continuum emission observed by ISO between 2 and 200~\mm, we used amorphous carbon (AC) dust (the nebula is carbon-rich, see MG09). We find a slightly better fit using BE1 type of amorphous carbon \citep{1991Rouleau_apj377}.

\begin{figure}
   \centering
 \includegraphics[width=9cm]{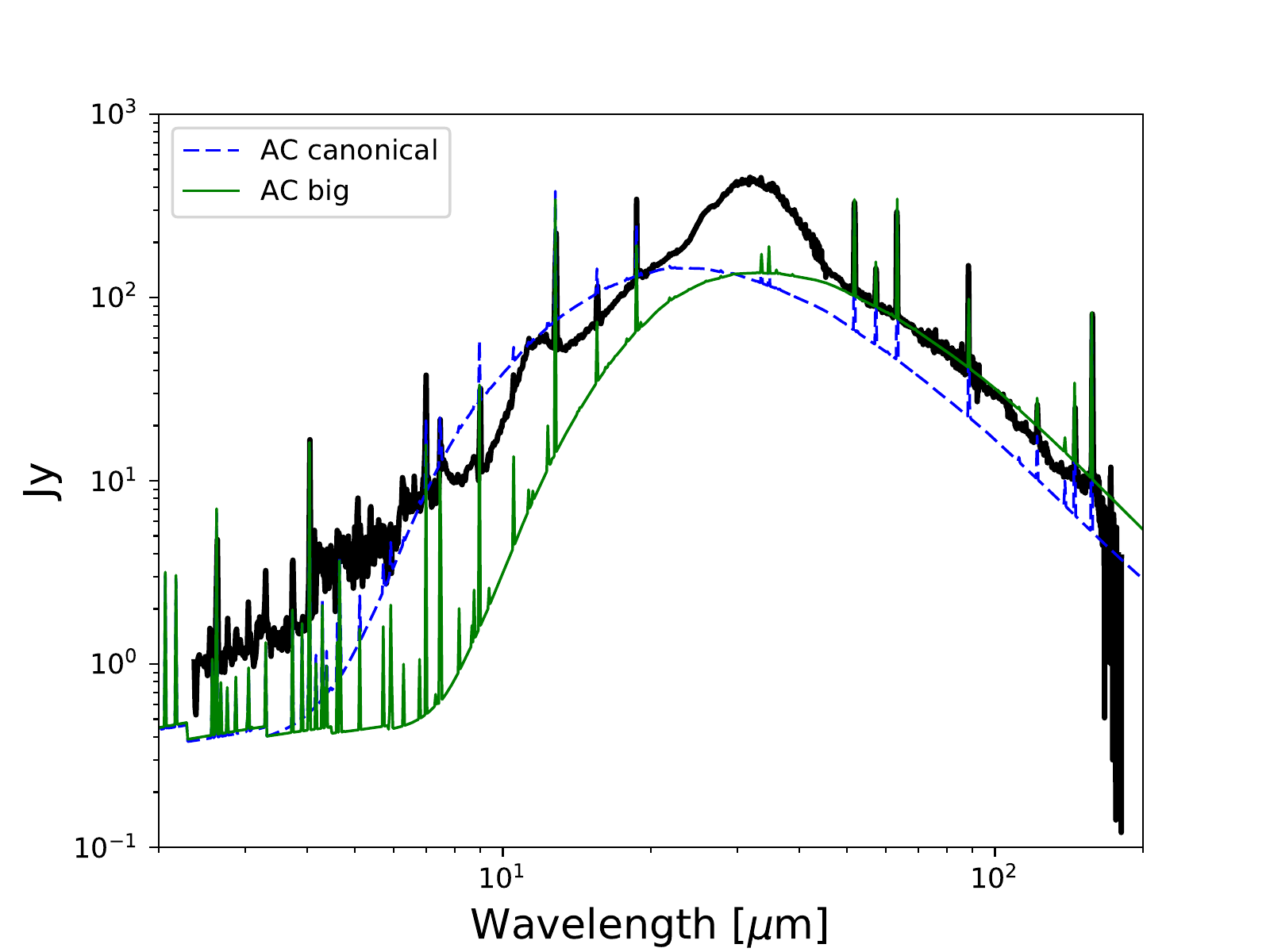}
   \caption{Dust emission of amorphous carbon. The blue dashed line represents  canonical size (0.005 to 0.25~\mm) grains in the whole nebula, and the green solid line represents big (0.03 and 0.4~\mm) grains only in the PDR. PN IC~418 observations are in black.}
              \label{fig:cont}%
\end{figure}

With dust of canonical sizes (from 0.005 to 0.25~\mm, following a -3.5 slope power law, and using ten different sizes of grains logarithmically distributed on the size range) distributed in the whole nebula (ionized region and PDR), the resulting emission exhibits a too-hot continuum (the peak emission is close to 20~\mm instead of the observed 30~\mm peak, see blue dashed line in Fig. \ref{fig:cont}). To reduce the dust temperature, one can put the dust in the outer part of the nebula, or increase its size (see Appendix \ref{append:degeneracy}). We chose to first adjust the far-IR observations (50-200~\mm) without over-predicting the observed emission at lower wavelengths: for this we used a size distribution of big grains between 0.03 and 0.4~\mm following 
power law size distribution, only in the PDR region. In our best fit model we find that a slope of -3.7 gives a better result than the classical one of -3.5.

The resulting dust emission (shown in green in Fig. \ref{fig:cont}) fits the 
long wavelengths observations, and is too low in the shorter wavelengths range. To solve this problem, we added other size and type dust grains, as described in the next sections.

\subsection{Adding very small grains}
\label{sec:adding-small-grains}
\begin{figure}
\centering
\includegraphics[width=9cm]{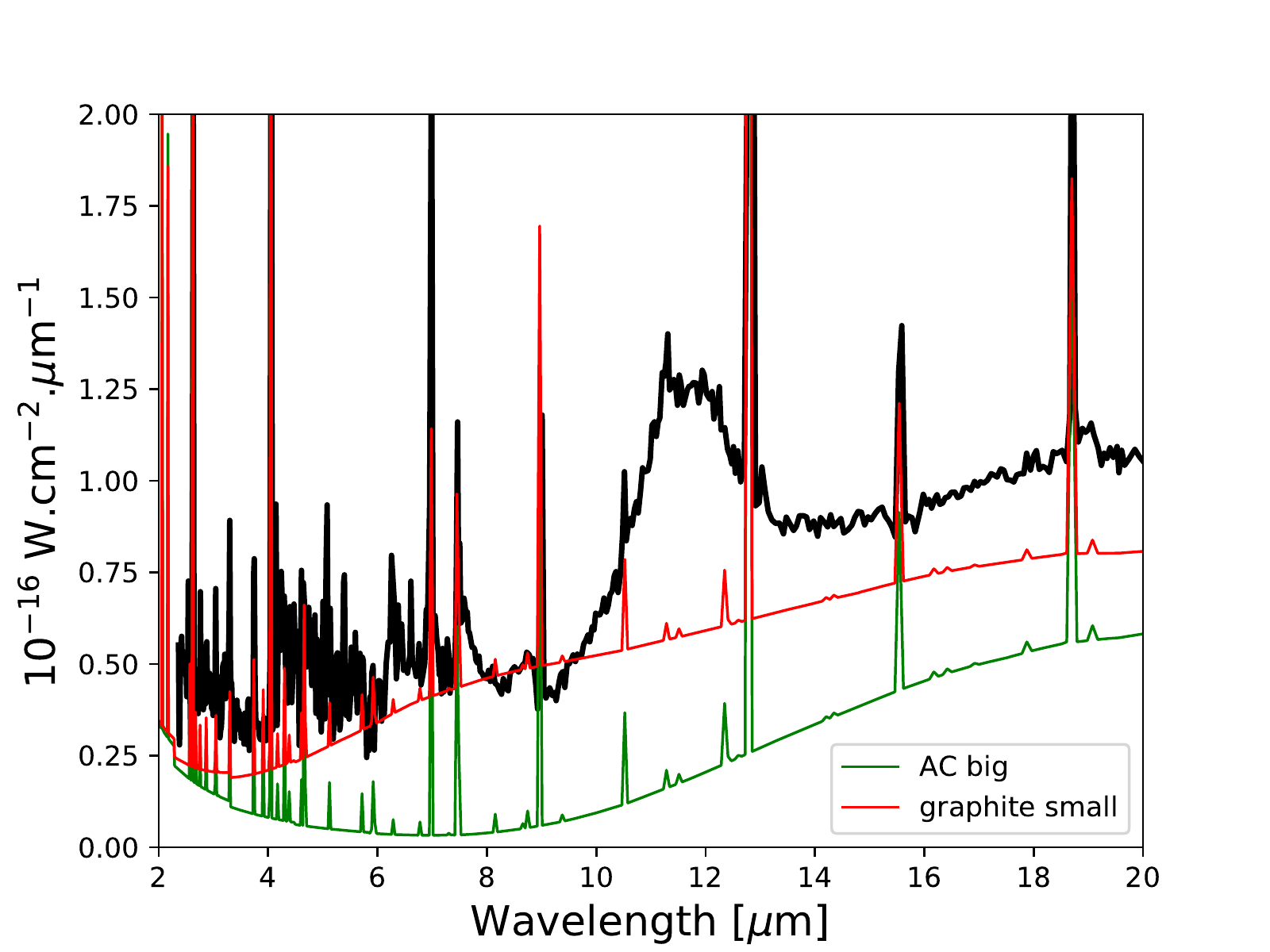}
   \caption{Observed ISO spectrum (black) and results of models with only big cool dust (green line), and big cool and additional smaller grains (red line).}
              \label{fig:small}%
\end{figure}

To increase the emission in the near-IR (2-10~\mm), we add a component of very small grains (sizes from 0.0005 to 0.03~\mm, distributed into ten grain sizes following a -3.5 slope power law) of graphite type. We obtain a better fit using graphite than using amorphous carbon. We put these small grains in the ionized region only, to increase the quality of the fit and following results on radiative pressure on dust grains by \citet{2018MNRAS.474.1935I}. Cloudy takes into account the quantum heating for these small grains.
The emission obtained adding these grains is shown in red in Fig.~\ref{fig:small}, compared to the emission from only big grains in the PDR (in green). The observed spectrum in black is clearly better fitted in the very short wavelengths range when this component is taken into account. Nevertheless, emission is still missing in the middle wavelength range (10 to 50 \mm), this will be fitted by SiC and MgS grains, as described in the next section.
The presence of such small grains close to the star seems to contradict the grain destruction by sputtering described by for example, \citet{1978Barlow_mnra183} or \citet{1979Draine_apj231} or more recently by \citet{2006Weingartner_apj645}. On the other hand, depletion of refractory elements, especially Fe, Si, and Ca \cite[see e.g.,][]{1983Shields_103, 1997Volk_180}, indicates that dust must still be mixed with the ionized gas. It is possible, that process of sputtering or shattering is not so efficient as thought, and the resulting small particles still remain in the ionized part of nebulae \cite[see e.g.,][]{1989Lenzuni_apj345}.

\section{Special features modeling}\label{sec:spec-feat-model}

Both 11.5~\mm and 30~\mm features are clearly observed in the IR spectrum of IC418. We explored the emission of SiC and MgS as usual suspects for these features.

\subsection{The 30 \mm band}
\label{sec:mgs}
For the 30~\mm broad feature, we adopt MgS spherical (Mie theory) and ellipsoidal grains \citep{1957vandeHulst,1983Bohren}. The optical constants are obtained from Jena database \footnote{\url{http://www.astro.uni-jena.de/Laboratory/OCDB/sulfides.html}.}.
This database offers optical n and k data from 10 to 50 \mm. Since no data are available at shorter wavelengths where absorption occurs, we choose to extrapolate them on the basis of the BE1 amorphous carbon data which are included in the Cloudy distribution \citep{1991Rouleau_apj377}. For larger wavelengths, we extrapolate a constant value for n, and an inverse wavelength law for k. 

The size distribution is following a -3.5 slope power law between 0.005 and 0.25~\mm. The grains are in the whole nebula. Following the case of the graphite and amorphous carbon described in the previous sections, we use ten different sizes to cover this size range.

\begin{figure}
\centering
\includegraphics[width=9cm]{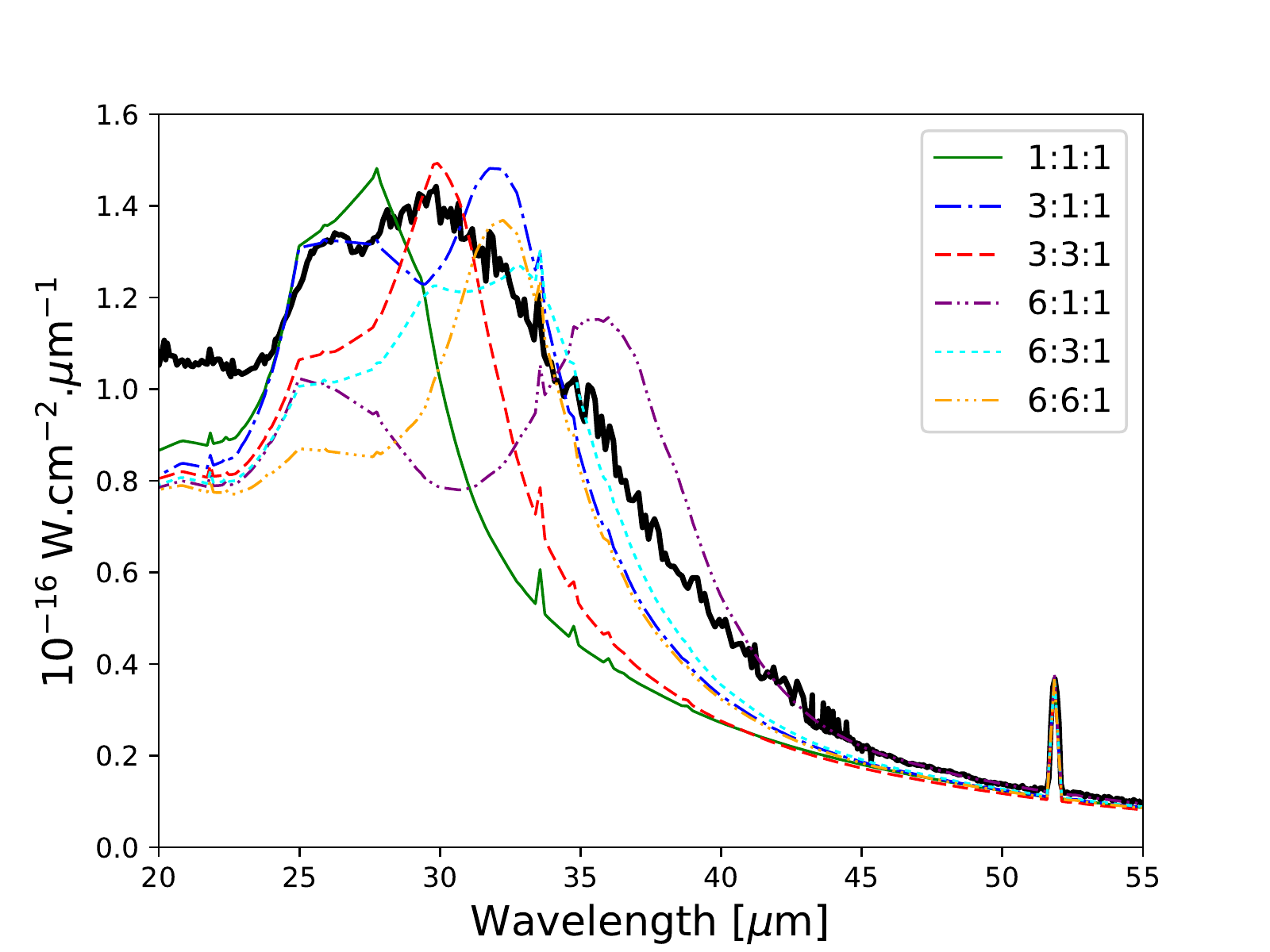}
   \caption{Emission of MgS grains of different shapes. IC~418 ISO observation is in black.}
              \label{fig:abc_MgS}%
\end{figure}

The Jena database offers data from Mg$_{0.9}$Fe$_{0.1}$S to FeS.
We explored the extreme values of these molecules and found that FeS does not lead to emission comparable to the observations. We used the Mg$_{0.9}$Fe$_{0.1}$S for the rest of the study.
The ellipsoidal approximation is obtained by applying \citet{2002Hony_aap390} procedure to convert spherical optical data into ellipsoidal ones, defined by their three semi-axes ratios a:b:c.

We explored spherical grains and five ellipsoidal morphologies corresponding to a:b:c = 3:1:1, 3:3:1, 6:1:1, 6:3:1, and 6:6:1. Figure \ref{fig:abc_MgS} shows the emission obtained for these grains. These results are very comparable to the ones obtained by \citet[][see their Fig. 8]{2002Hony_aap390}. The main effect of changing the shape of the grains is shifting the emission to larger wavelengths compared to spherical case (green line in Fig.~\ref{fig:abc_MgS}). 
None of these single morphology emissions can reproduce the observation. A mixture of these morphologies is then used by running models where different contribution of each shape to the total amount of grain are considered. We converged a grid of models to the best fit, obtained by minimizing a $\chi^2$ between 25 and 40 \mm.
We note that, contrary to the case of featureless emission like the one of amorphous carbon for which the grain temperature largely determine the location of the emission peak, the emission of MgS (and SiC) is not significantly changed when the temperature changes.

Using this combination of ellipsoidal grains leads to a better fit to the observations. The best model presented in Fig.\ref{fig:best} (see Sect.~\ref{sec:best}) shows in lower-right panel the fit of the 30~\mm feature. We see that 30 micron feature is reproduced fairly well. 

\subsection{The 11.5 \mm band}
\label{sec:sic}

\begin{figure}
\centering
\includegraphics[width=9cm]{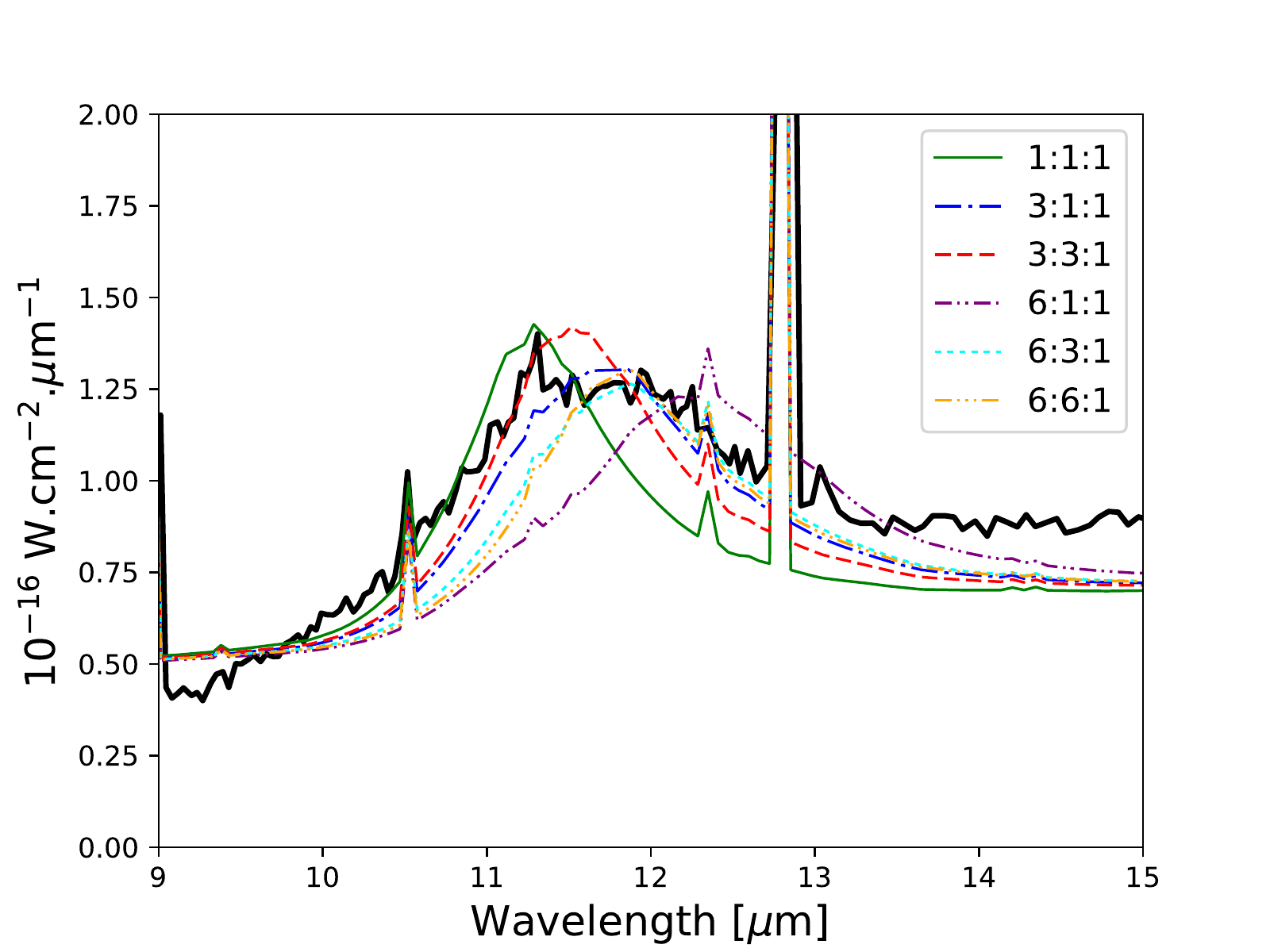}
   \caption{Emission of SiC grains of different shapes. IC~418 ISO observation is in black.}
              \label{fig:abc_SiC}%
\end{figure}

The 11.5 \mm band has also been observed by the spectrograph on board the Spitzer Space Telescope, see Figure 1 from \citet{2018AJ....155..105D}. A correction factor is needed to make both Spitzer and ISO spectra match more or less, but in both observations the structure of the 11.5 \mm band looks the same.

SiC emission is obtained by including dielectric data from \citet{1988Pegourie_aap194} in Cloudy. These data are between 0.1~\mm and 243~\mm. At shorter wavelengths, for n we extrapolated on the basis of the BE1 amorphous carbon from Cloudy distribution \citep{1991Rouleau_apj377} and for k we extrapolate using a power law of slope 2. For longer wavelengths, for n we extrapolate a constant value and for k a power law of slope -1.

We used the same method as for MgS to obtain the optical properties for ellipsoidal grains. 
The resulting emission is shown in Fig.~\ref{fig:abc_SiC} with the observed spectrum. 
The canonical size distribution is following a -3.5 slope power law between 0.005~\mm and 0.25~\mm. The grains are in the whole nebula. We used ten different sizes to cover this size range.

The best fit, obtained by minimizing a $\chi^2$ between 9 and 16~\mm is presented in Fig.~\ref{fig:best}, center-right panel (see Sect.~\ref{sec:best}).
The fit is still not perfect and we may try other optical properties when available or other type of grains.

\section{Best model}
\label{sec:best}

\begin{figure}
\centering
\includegraphics[width=9cm]{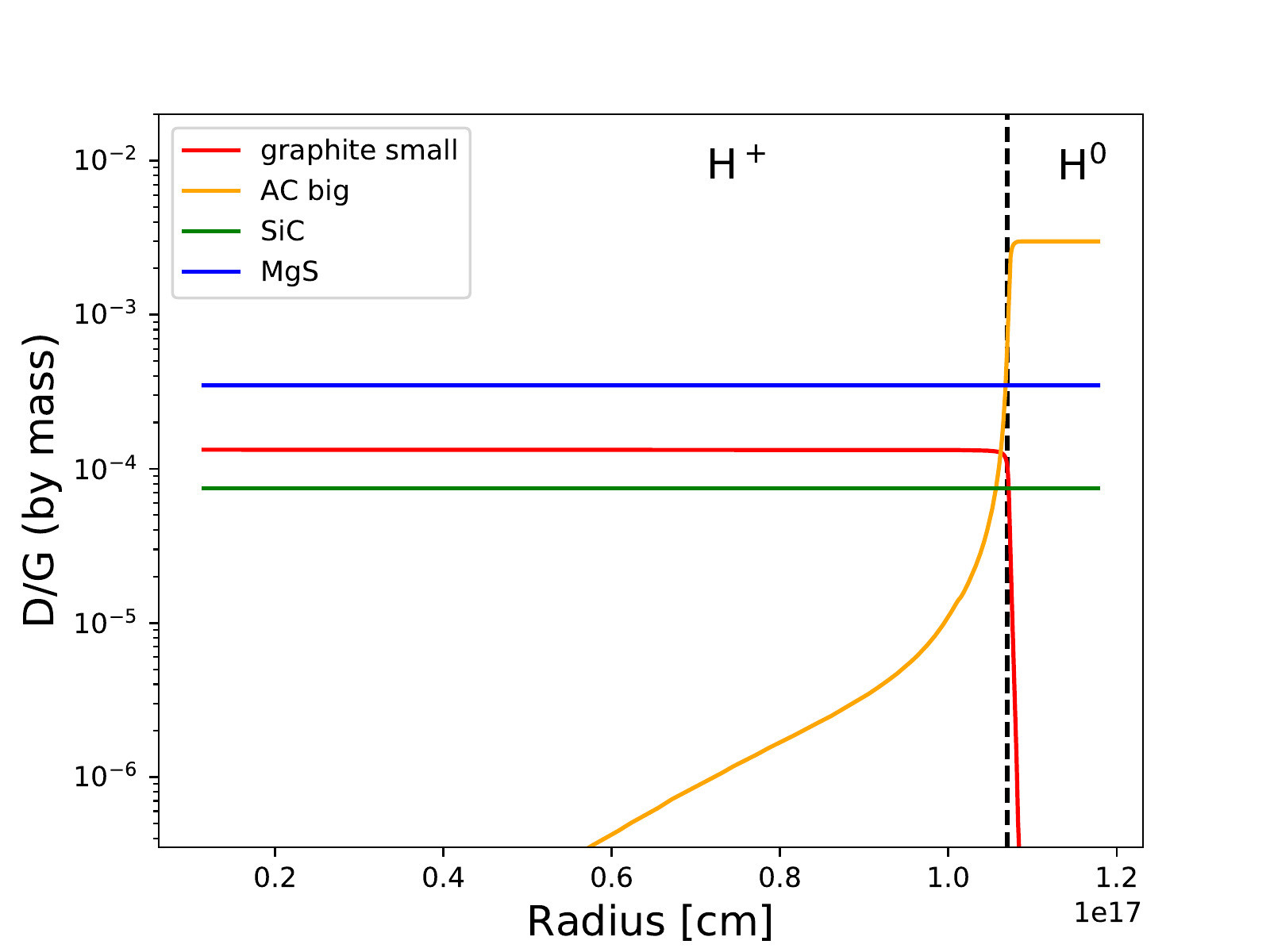}
\caption{D/G by mass for the different types of dust grains. The vertical dashed line delimits the H$^+$ region and the PDR. 
For the SiC and MgS dust types the D/G considers the sum of D/Gs for all the shapes for the best model (5 ellipsoidal and 1 spherical, see Table~\ref{tab:DG}).} 
\label{fig:dgratio}
\end{figure}

\begin{figure}
\centering
\includegraphics[width=9cm]{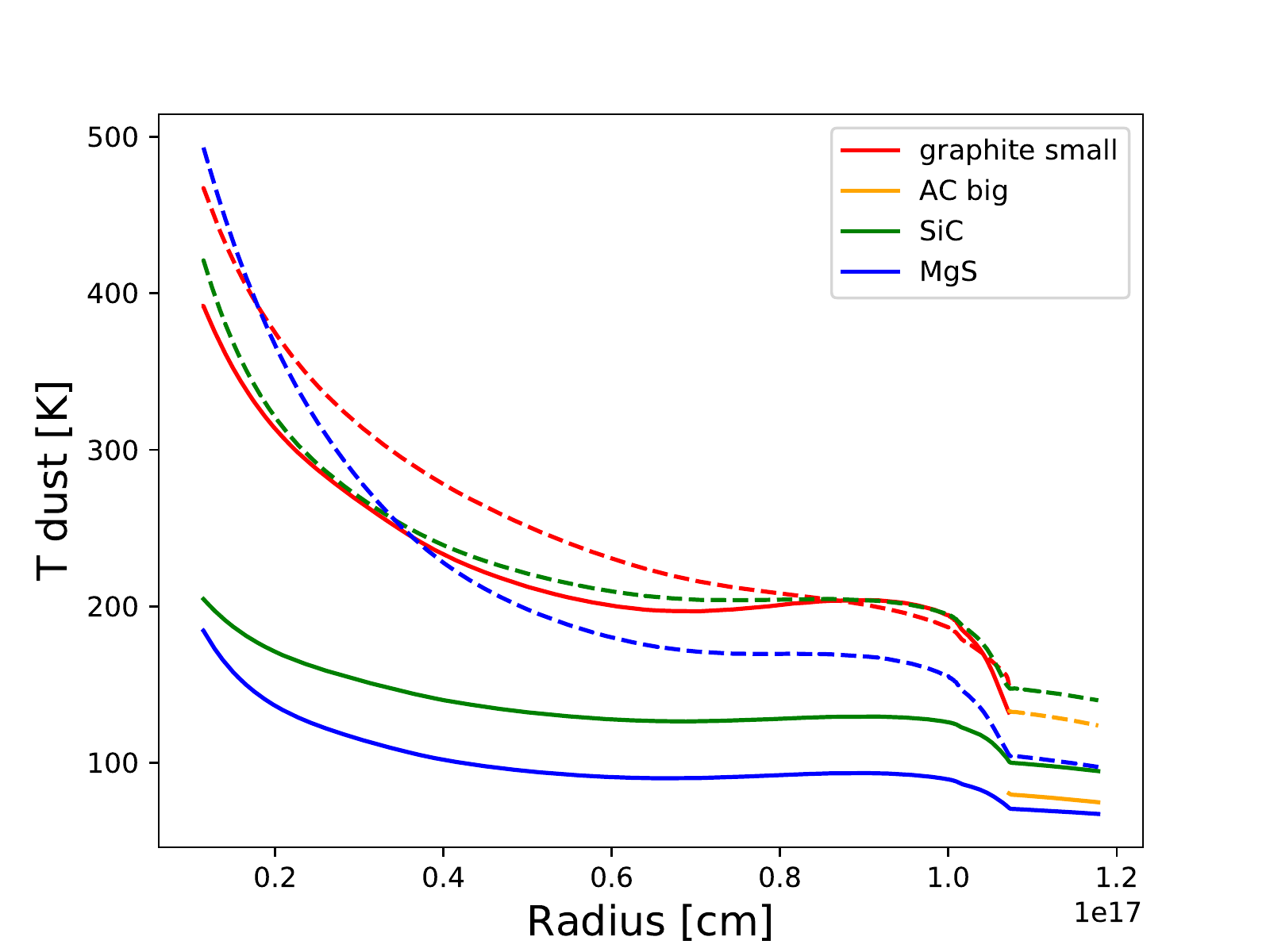}
\caption{Dust temperature for different grain sizes. Each color corresponds to different grain types: green and blue represent canonical distribution of dust sizes in case of SiC and MgS, while red and yellow represent small and big size distributions, respectively.  Graphite and amorphous carbon grains are only present in H$^+$ and PDR regions, respectively. For each dust type, solid and dashed lines represent the biggest and smallest size in each grain size distribution. 
Only spherical SiC and MgS grains are shown.}
\label{fig:tdust}%
\end{figure}

\begin{figure*}
\centering
\includegraphics[width=16cm]{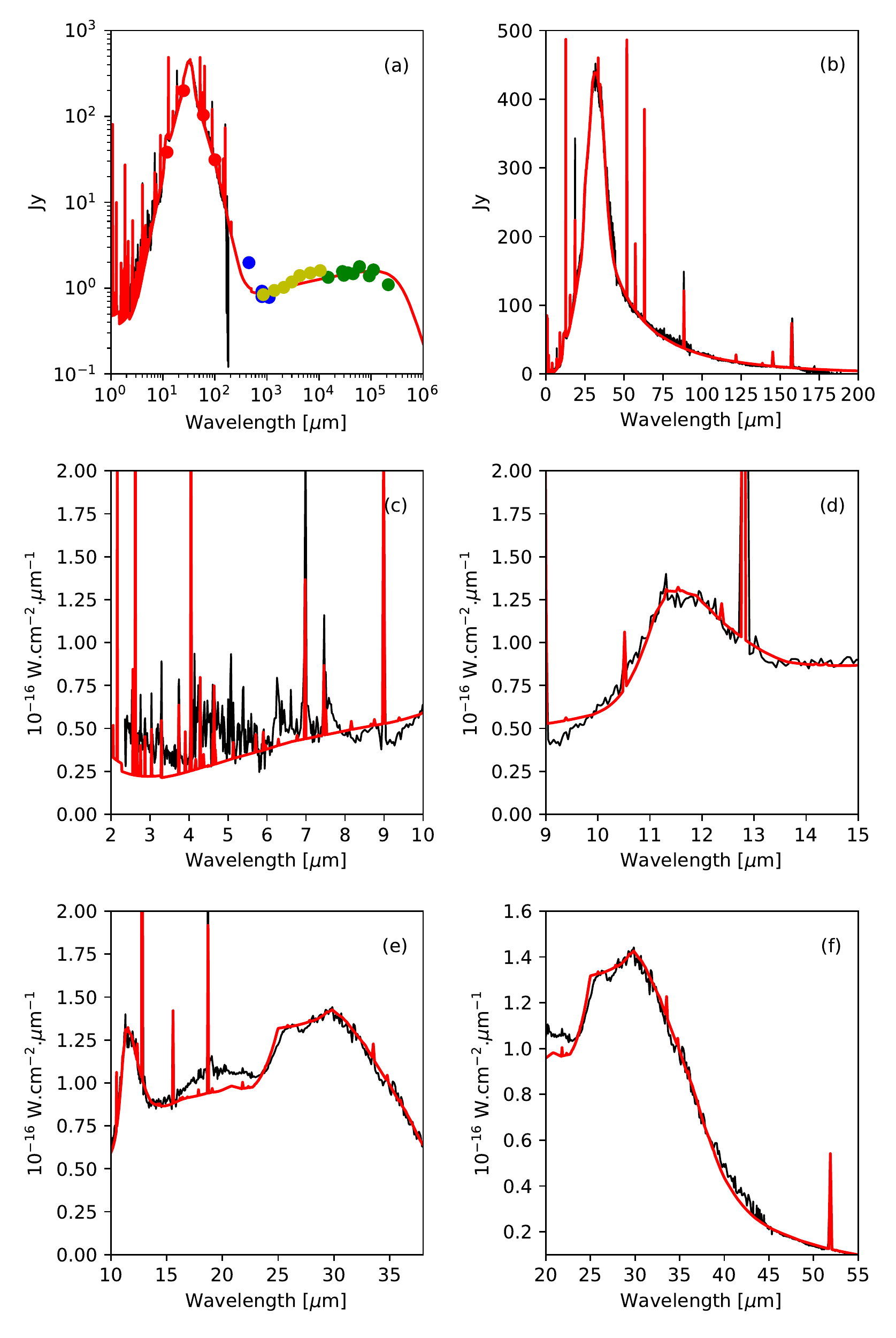}
\caption{Comparison between our best model and the observations. ISO data are black solid line, the model is red solid line. In the upper left panel, red dots show IRAS data, blue dots show data reported by \citet{1992Hoare_mnra258}, yellow dots show data from \citet{2015Planck-Collaboration_aap573}, and green dots show radio observations (see Sect.~\ref{sec:observations}). }
\label{fig:best}%
\end{figure*}
 
Some basic properties of our best model for each region (H$^+$ and PDR) are given in Table~\ref{tab:results}.
Our best model is obtained by adding four dust types (namely amorphous carbon, graphite, SiC and MgS) of different size distributions, shapes, and even position in the nebula. All the grains are distributed into ten size bins each. Additionally SiC and MgS grains use six shapes each. This finally leads to 140 different grains included in the Cloudy model.  
All these properties are summarized in Table \ref{tab:DG}, where the dust-to-gas ratio by mass are given for each grain type.

\begin{table}[h]
\centering
\begin{tabular}{l l l  }
\hline \hline
    & H$^+$ region & PDR \T\B\\
\hline \hline
Volume [10$^{51}$ cm$^3$] & 5.1 & 1.5 \T \\
Inner radius [10$^{17}$ cm] & 0.115 & 1.085$^*$\\
Outer radius [10$^{17}$ cm] & 1.068$^*$ & 1.178\\
filling factor & 1.00 & 0.10 \\
H-Mass [M$_\odot$] & 0.047 & 0.040 \\
D/G by mass [10$^{-4}$]& 5.56 & 34.1 \\ 
$<n_H>$ [cm$^{-3}$] & 1.1$\times$10$^4$ & 3.0$\times$10$^5$ \\
$<T_e>$ [K]& 9031 & 462 \B \\
\hline \B
\end{tabular}
\caption{Properties of the H$^+$ region and the PDR. 
$^*$The transition region between the outer radius of the H$^+$ region and the inner radius of the PDR only contributes to 0.1\% of the total H-Mass and is not considered in this table.}
\label{tab:results}
\end{table}

\begin{figure}
\centering
\includegraphics[width=9cm]{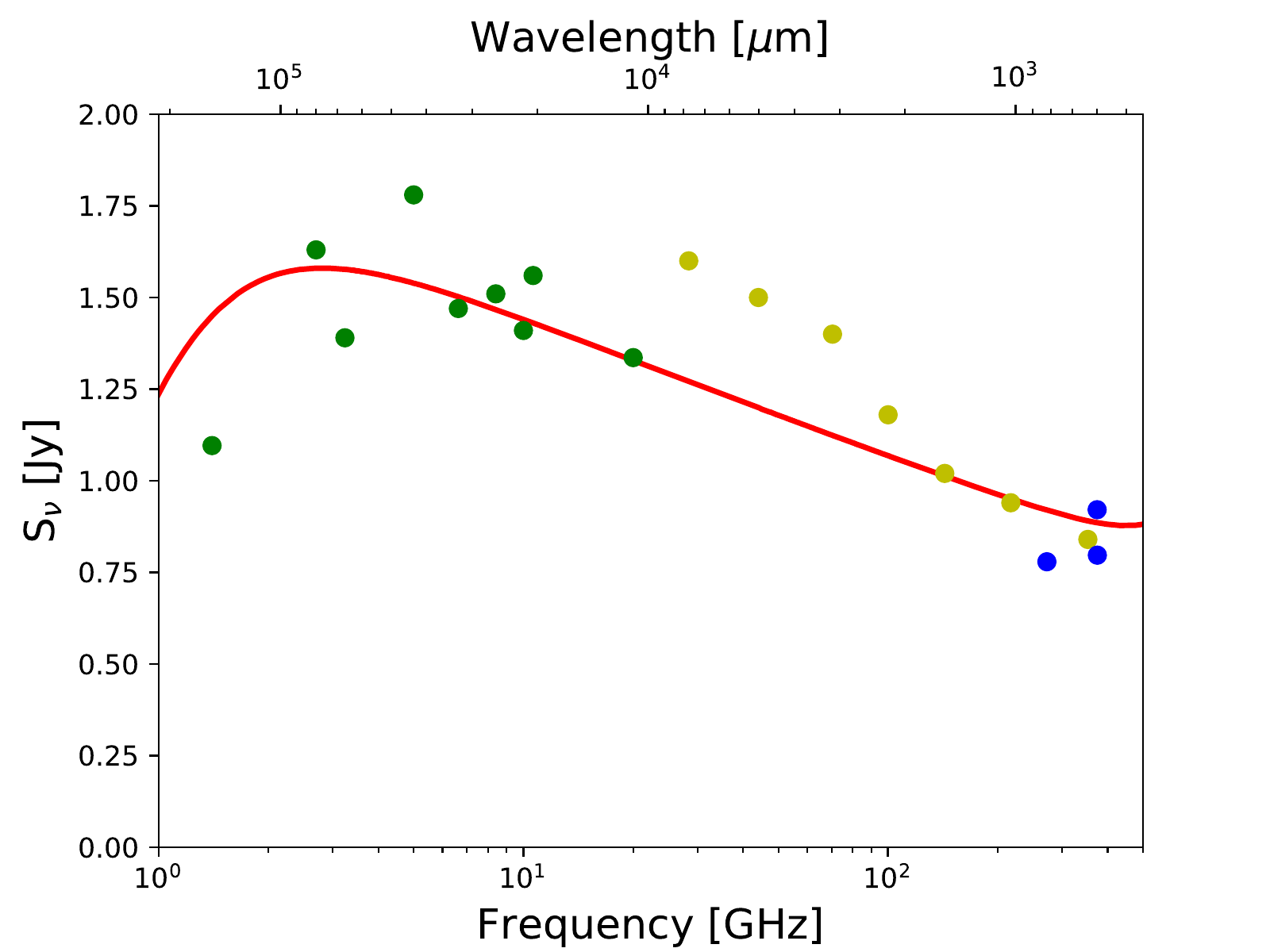}
\caption{Red solid line is the fit of the model at longer wavelengths. The dots are observations from different authors: green dots are radio observations (see Sect. \ref{sec:observations}), yellow dots are data from \citet{2015Planck-Collaboration_aap573} and blue dots are data reported by \citet{1992Hoare_mnra258}.}
\label{fig:radio}%
\end{figure}

\begin{table*}
\centering
\begin{tabular}{l c c c c c c}
\hline \hline
Grain type & D/G by mass$^*$& $a_{min}$(\mm) & $a_{max}$(\mm)&P.L. slope  & H$^+$ & H$^0$ \T\B \\
\hline \hline
Graphite$^{**}$ & 1.32$\times 10^{-4}$ &  0.0005 & 0.03 & -3.5& X  &   \T \\ 
AC$^{**}$ & 2.60 $\times 10^{-3}$ & 0.02 & 0.4& -3.7 &  & X \\
SiC (1:1:1 \& 6:1:1) &  2.25$\times 10^{-5}$ &  0.005 & 0.25& -3.5 & X & X\\
SiC (remaining shapes) & 7.50$\times 10^{-6}$ & 0.005 & 0.25 & -3.5& X & X\\
MgS (1:1:1 \& 3:1:1 \& 6:1:1) & 3.32$\times 10^{-5}$ & 0.005 & 0.25& -3.5 & X & X  \\
MgS (3:3:1) & 8.29$\times 10^{-5}$ & 0.005 & 0.25 & -3.5& X & X  \\
MgS (6:1:1) &  9.95$\times 10^{-5}$ & 0.005 & 0.25 & -3.5& X & X \\
MgS (6:3:1) &  6.63$\times 10^{-5}$ & 0.005 & 0.25 & -3.5& X & X  \B \\
\hline
\multicolumn{5}{l}{
  \begin{minipage}{10.cm}~\\
$^{*}$ For each dust type and shape.\\ 
$^{**}$ D/G considering only the region where the dust is present.
  \end{minipage}
}\\
\end{tabular}

\caption{Properties of all the dust types and shapes of the grains used in our best model.
Dust to gas ratio by mass for each dust type and shape are given in column 2. The lower and upper values for the grains sizes are given in columns 3 and 4, respectively. Columns 5 and 6 indicate with a cross where the dust grains are present.}
\label{tab:DG}
\end{table*}

In Figure \ref{fig:dgratio} we plot the D/G  (by mass) for the different grains types. While some of them are constant over the whole nebula, others are only in one part: big amorphous carbon grains are mostly in the H$^0$ (external) region (see Appendix A for details of their distribution), and small graphite grains in the H$^+$ (inner) region. MgS and SiC grains (see previous sections) are in the whole nebula. For these grains, we plot the sum of D/G's for all the shapes.
Molar masses for C, Mg$_{0.9}$Fe$_{0.1}$S and SiC are 12.0, 59.5 and 40.1, respectively.
The density for MgS is 2.7 g/cm$^3$ \citep{ropp2012encyclopedia}; we then use 3.05~g/cm$^3$ for Mg$_{0.9}$Fe$_{0.1}$S. For SiC we use 3.21~g/cm$^3$ \citep{patnaik2003handbook}.

In Figure \ref{fig:tdust} we show the variation of the dust temperature among the radius for different grain sizes and types, biggest grains (solid lines) being cooler. In the case of SiC and MgS, the effect of the shape on the temperature (not shown here) is significantly smaller than the effect of the size, for a given grain type.

In the Fig.~\ref{fig:best} we present the comparison between our best model and the observations, for various wavelength ranges and using various units for the fluxes. The emission lines are not aimed to be compared in these figures, as the wavelengths meshes are not comparable. Emission lines in the Cloudy output appears most of the time on only one pixel, leading to triangle profiles. We concentrated here on the dust continuum emission. 
The (a) and (b) upper panels show the whole fit using Jy unit, from 1~\mm to 1~m (in log scale) and from 2~\mm to 200~\mm in linear scale, respectively. Some observations obtained through spectral bands (see caption) are also plotted in panel (a). 
In Figure~\ref{fig:radio} we present 
enlargement of panel (a), in which we compare the observation and the model fluxes for the radio wavelength range.
The total energy emitted by the nebula is 1.13~10$^{37}$ erg/s, of which almost 65\% is emitted 
above 2.5~\mm\ (42\% from the H$^+$ region and 23\% from the PDR) leading to a total of 7.4~10$^{36}$ erg/s, including the small contribution from the emission lines of $\sim$~2~10$^{35}$ erg/s.
The other panels show close-ups at some wavelength ranges: panel (c) is showing the 2-10~\mm range, where the small grains mainly emits; the panel (d), (e), and (f) show the SiC and MgS emissions. As we can see from Fig.~\ref{fig:best}, the best fit reasonably well reproduces the infrared continuum and the main dust features at 11.5 and 30 \mm, with clear discrepancy between 12 and 25 \mm. This discrepancy suggests that some specific dust component was not taken into account during modeling, and is discussed in Sect.~\ref{sec:discussion}.
As a final result, we report in Table~\ref{table:abund} the abundances of some elements in the gaseous phase and in the dust, as well as the total value and the solar values. 
\begin{table}
\centering                                      % used for centering table
\begin{tabular}{l l l l l}          % centered columns (4 columns)
\hline\hline                        % inserts double horizontal lines
Element  & Gaseous phase & Dust & Total  & Solar$^*$ \T\B \\    % table heading
\hline\hline                                % inserts double horizontal
C  & -3.15 & -4.71 & -3.14 & -3.57 \T \\
Mg & -5.07$^{**}$ & -5.10 & -4.78 & -4.40 \\
Si & -5.55$^{***}$ & -5.55 & -5.25 & -4.49 \\
S  & -5.35 & -5.05 & -4.87 & -4.88 \\
Fe & -7.4 & -6.05 & -6.03 & -4.50 \B \\
\hline                                             %inserts single line
\multicolumn{5}{l}{
  \begin{minipage}{8.cm}~\\
$^*$ Solar values from \citet{2009Asplund_araa47}.

$^{**}$ This value differs slightly from the one obtained by MG09 because of changes in atomic data between different versions of Cloudy.

$^{***}$ This value is obtained by fitting the \forbm{Si}{ii}{34.8} line, mainly emitted by the PDR.
  \end{minipage}
}\\
\end{tabular}
\caption{Abundances (by number, in log) of some elements in PN IC418. For carbon, only the H$^+$ is considered (graphite grains, see Tab\ref{tab:DG}).}              % title of Table
\label{table:abund}      % is used to refer this table in the text

\end{table}

\begin{figure}
\centering
\includegraphics[width=9cm]{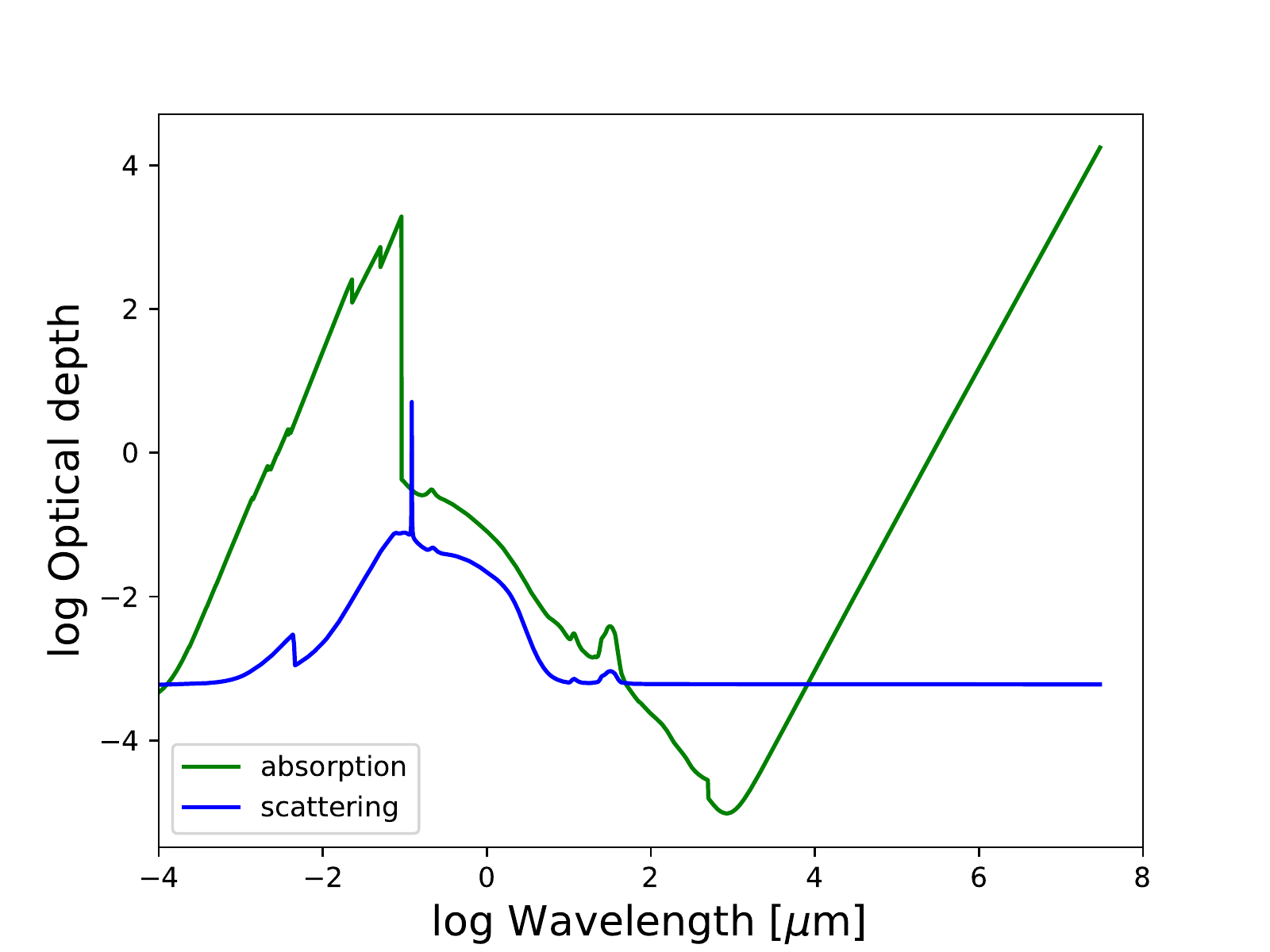}
\caption{Optical depth of the best model over a spectral range from X-rays to Radio. Effect of scattering and absorption are separated.}
\label{fig:optdepth}
\end{figure}

In Figure~\ref{fig:optdepth} we show the optical depth of the nebula relative to the wavelength, ranging from X-rays to radio. We can see that neither the scattering nor the absorption are close to be optically thick in the IR range we are modeling in this work.
Thus, we are convinced that radiative transfer in CLOUDY gives reasonable results.

\section{Discussion}\label{sec:discussion}

The model presented in the previous section as our best fit suffers some defects coming from the simplifications 
that could be improved in a future work. For example, the broad
30\,\mm feature, 
could perhaps be better reproduced using grains with SiC core coated with MgS mantles as described by \citet{2008Zhukovska_aap486}. 
We tried FeS and MgS without any success in reproducing the observed spectrum.

We were able to compare the gaseous and total abundances, and obtain with the ones derived by \citet{2017Dopita_mnra470}. We found comparable values for all the elements in the gaseous phase (except for Si, which is determined from a line emitted by the PDR, not taken into account by \citeauthor{2017Dopita_mnra470}). On the other side, the difference is very important when comparing the abundances in the dust phase: the values obtained by \citeauthor{2017Dopita_mnra470} are systematically higher than ours (except for S/H). The values given by \citeauthor{2017Dopita_mnra470} are not obtained by fitting the IR emission, but rather by applying the dust depletion factor determined by \citet{2009Jenkins_apj700}, normalized by the Fe/H depletion. The depletion of iron in IC418 is quite puzzling, and should perhaps not be taken as reference for the other elements; it is the most important depletion observed for iron in PNe \citep{2016Delgado-Inglada_mnra456}.
As an illustration, our total value for log(Mg/H) is -4.78, while it is -3.68 in \citeauthor{2017Dopita_mnra470}, values that may be compared to the solar value of -4.47. Our results have obvious limitations, but the total abundances we obtain are at least lower or close to the solar values (except for C/H, which is mainly in the gaseous phase and only constraint in the ionized region). 

Our main goal for this work was to explain the depletion of Mg, Si, S, and Fe in the ionized nebula by their presence in form of dust grains. The model presented here, which reproduces the global energy budget of the IR dust emission, and therefore should involve the right amount of dust, still predicts too low abundances of these elements compared to the solar value (which is a valid reference for IC418). We actually found that, for example, the amount of Mg and Fe embedded in Mg$_x$Fe$_{1-x}$S is not enough to reach the solar value, see Table~\ref{table:abund}.

Some under-abundant elements may be incorporated into missing dust component (see discussion of Fig.~\ref{fig:om_ratio} below). One the other side, S/H is solar once adding the gaseous and dusty phases. This is in clear contradiction with what is found by \citet{2009Zhang_apj702}, who argue that the amount of MgS, (estimated from the available sulfur), which is necessary to reproduce the IR emission on the proto-PN \object{HD 56126} is much higher than the amount available for this object. 

We must mention here that we did not have taken into account the effects of coating. The coating can hidden elements inside grains that do not show their presence by the emission defined by their surface composition. It also increases the apparent element abundance of an element that is only present on the surface of the grains. The coating may also be different between the ionized region and the PDR.

In Fig.~\ref{fig:om_ratio} we show the residues between our best fit and the observations, between 9 and 37~\mm. The feature drawn here is well above the noise and can be due to our poor knowledge of the optical properties of the MgS and SiC emissions, but it may also correspond to an unknown carrier. 

The very broad feature between 15-25 \mm could be due to some kind of solid with a mixed aromatic and aliphatic structure.
Similar broad emission plateaus observed in C-rich dust PNe like IC 418, for example, at $\sim$8 and 12~\mm, have been suggested to emerge from mixed aromatic-aliphatic organic nanoparticles (the MAON model proposed by \citet{2011Kwok_nat}). Such an organic component is missing in our model and its inclusion could affect the results of our IR modeling. In the MAON model, the mixed aromatic-aliphatic grains may be very diverse and complex (e.g., even including other atoms, apart from C and H, such as N, O, and S). For example, laboratory spectra of some hydrogenated amorphous carbon (HAC) samples, being just a particular case of these mixed aromatic-aliphatic MAON grains display broad emission features at 21 and 30 $\mu$m \citep[e.g.,][]{2001Grishko_apjl558}. Such a dust component could explain also the formation of complex dehydrogenated organics like fullerenes \citep[e.g.,][]{2010Garcia-Hernandez_apjl724, 2012Garcia-Hernandez_apj760} that are known to be present in IC~418 and other C-rich PNe \citep[e.g.,][and references therein]{2014Otsuka_mnra437}. Unfortunately, we do not have the optical constants of these MAON-like particles, except for a few specific HAC samples (W. Duley private communication). Preliminary modeling using this limited HAC data does not show the broad $\sim$12-25 $\mu$m feature observed in IC 418 (Morisset et al., in preparation).

\begin{figure}
\centering
\includegraphics[width=9cm]{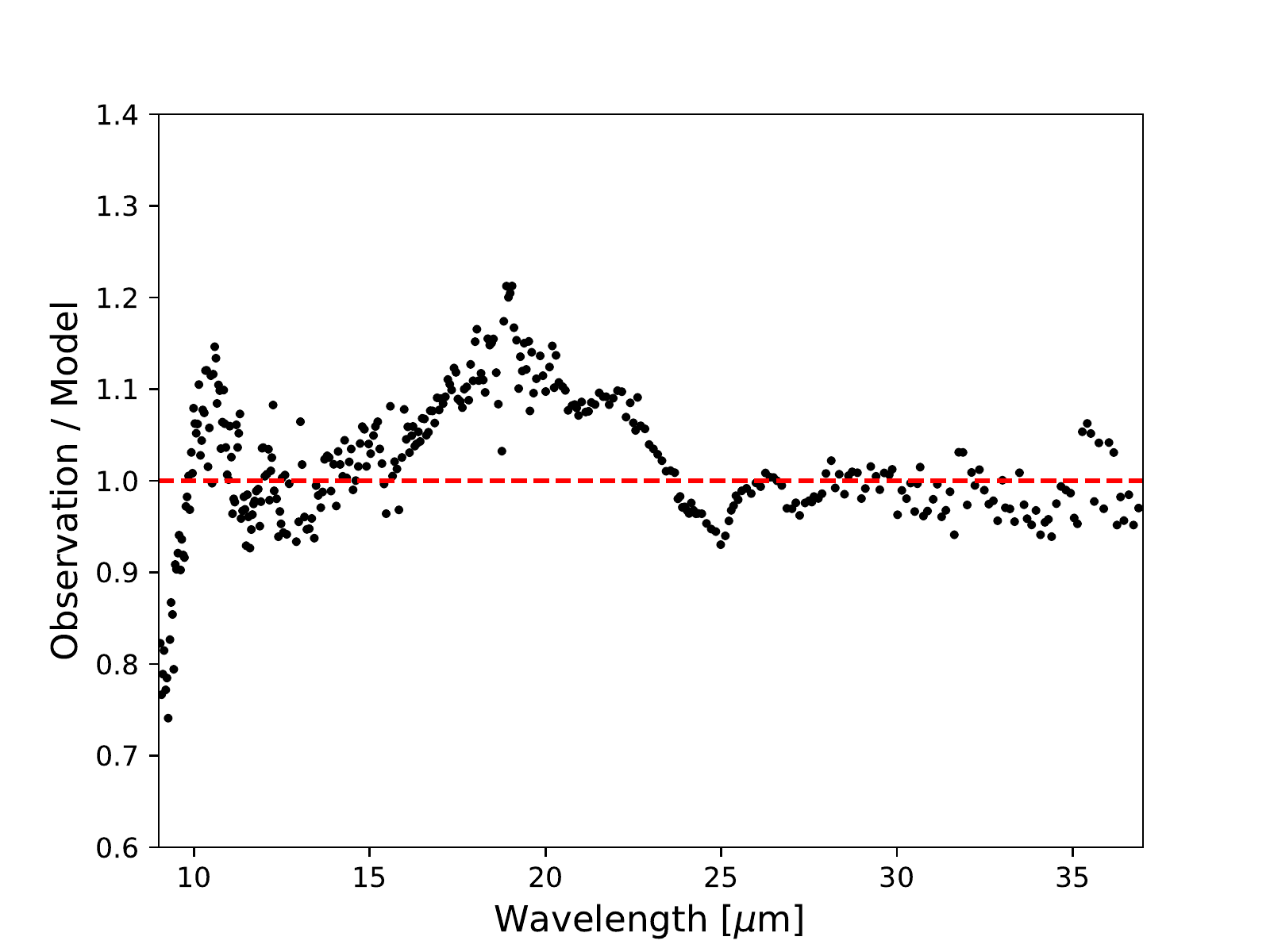}
\caption{Ratio between observation and our best model for wavelengths from 9 to 37~\mm.}
\label{fig:om_ratio}%
\end{figure}

\section{Conclusions}\label{sec:conclusions}

This is the first time, to our knowledge, that a full ionized region + PDR model of a PN is presented that fits the IR emission from 2 to 200~\mm, including the special features observed at 11.5 and 30~\mm (the very detailed and complete model of NGC~6781 presented by \citet{2017Otsuka_aaps231} does not include these features since they are not detected in that object). We succeed in reproducing the whole budget as well as those two broad emission features, with graphite, amorphous carbon, SiC and MgS grains. The observed emission at short wavelengths (2 to 9~\mm) is reproduced by small graphite grains located in the H$^+$ region, while the long wavelength emission (50 to 200~\mm) is due to carbon big grains located in the H$^0$ (also called PDR) region. Adding the contribution from these dust grains to the abundances of the elements already present in the gaseous form allow us to determine their total abundances. 

Our IR modeling presented here (i.e., the nature of the different dust components) might be proved or disproved by upcoming space-based IR facilities such as the James Webb Space Telescope (JWST). For example, high-spatial resolution imaging with the MIRI instrument on board JWST, may provide the relative spatial distribution of all these dust
species, to be compared with the spatial distribution found by our IR modeling. All the files needed to run the model presented in this work as well as the outputs of the Cloudy model are available on the web site \url{https://github.com/Morisset/IC418}.

\begin{acknowledgements}
This work is supported by grants DGAPA/PAPIIT-107215 and CONACyT-CB2015-254132. V.G-L received support from CONACyT graduated students grants, and from PAEP-DGAPA program from UNAM. R.Sz. acknowledges support from the Polish NCN grant 2011/01/B/ST9/02031. D.A.G.H. acknowledges support provided by the Spanish Ministry of Economy and Competitiveness (MINECO) under grant AYA$-$2014$-$58082-P.

We thanks Peter van Hoof, who helped on running complex Cloudy dusty models and made some comments on the draft. Discussions with Peter van Hoof and Gary Ferland helped a lot to clarify the limitations of Cloudy.

We thank the referee for the useful comments we received, improving the quality of the paper.

\end{acknowledgements}

\bibliographystyle{aa}
\bibliography{BIBLIO_min}

\begin{thebibliography}{56}
\expandafter\ifx\csname natexlab\endcsname\relax\def\natexlab#1{#1}\fi

\bibitem[{{Abrahamsson} {et~al.}(2007){Abrahamsson}, {Krems}, \&
  {Dalgarno}}]{2007Abrahamsson_apj654}
{Abrahamsson}, E., {Krems}, R.~V., \& {Dalgarno}, A. 2007, \apj, 654, 1171

\bibitem[{{Asplund} {et~al.}(2009){Asplund}, {Grevesse}, {Sauval}, \&
  {Scott}}]{2009Asplund_araa47}
{Asplund}, M., {Grevesse}, N., {Sauval}, A.~J., \& {Scott}, P. 2009, \araa, 47,
  481

\bibitem[{{Barlow}(1978)}]{1978Barlow_mnra183}
{Barlow}, M.~J. 1978, \mnras, 183, 367

\bibitem[{Bohren \& Huffman(1983)}]{1983Bohren}
Bohren, C.~F. \& Huffman, D.~R. 1983, Absorption and Scattering of Light by
  Small Particles ({WILEY-VCH Verlag GmbH})

\bibitem[{{Cahn} {et~al.}(1992){Cahn}, {Kaler}, \&
  {Stanghellini}}]{1992Cahn_aaps94}
{Cahn}, J.~H., {Kaler}, J.~B., \& {Stanghellini}, L. 1992, \aaps, 94, 399

\bibitem[{{Dayal} \& {Bieging}(1996)}]{1996Dayal_apj472}
{Dayal}, A. \& {Bieging}, J.~H. 1996, \apj, 472, 703

\bibitem[{{Delgado-Inglada} {et~al.}(2016){Delgado-Inglada}, {Mesa-Delgado},
  {Garc{\'{\i}}a-Rojas}, {Rodr{\'{\i}}guez}, \&
  {Esteban}}]{2016Delgado-Inglada_mnra456}
{Delgado-Inglada}, G., {Mesa-Delgado}, A., {Garc{\'{\i}}a-Rojas}, J.,
  {Rodr{\'{\i}}guez}, M., \& {Esteban}, C. 2016, \mnras, 456, 3855

\bibitem[{{D{\'{\i}}az-Luis} {et~al.}(2018){D{\'{\i}}az-Luis},
  {Garc{\'{\i}}a-Hern{\'a}ndez}, {Manchado}, {Garc{\'{\i}}a-Lario}, {Villaver},
  \& {Garc{\'{\i}}a-Segura}}]{2018AJ....155..105D}
{D{\'{\i}}az-Luis}, J.~J., {Garc{\'{\i}}a-Hern{\'a}ndez}, D.~A., {Manchado},
  A., {et~al.} 2018, \aj, 155, 105

\bibitem[{{Dixon}(1970)}]{1970Dixon20}
{Dixon}, R.~S. 1970, \apjs, 20, 1

\bibitem[{{Dopita} {et~al.}(2017){Dopita}, {Ali}, {Sutherland}, {Nicholls}, \&
  {Amer}}]{2017Dopita_mnra470}
{Dopita}, M.~A., {Ali}, A., {Sutherland}, R.~S., {Nicholls}, D.~C., \& {Amer},
  M.~A. 2017, \mnras, 470, 839

\bibitem[{{Dopita} {et~al.}(2005){Dopita}, {Groves}, {Fischera}, {Sutherland},
  {Tuffs}, {Popescu}, {Kewley}, {Reuland}, \& {Leitherer}}]{2005Dopita_apj619}
{Dopita}, M.~A., {Groves}, B.~A., {Fischera}, J., {et~al.} 2005, \apj, 619, 755

\bibitem[{{Draine} \& {Salpeter}(1979)}]{1979Draine_apj231}
{Draine}, B.~T. \& {Salpeter}, E.~E. 1979, \apj, 231, 438

\bibitem[{{Ferland} {et~al.}(2017){Ferland}, {Chatzikos}, {Guzm{\'a}n},
  {Lykins}, {van Hoof}, {Williams}, {Abel}, {Badnell}, {Keenan}, {Porter}, \&
  {Stancil}}]{2017Ferland_rmxaa53}
{Ferland}, G.~J., {Chatzikos}, M., {Guzm{\'a}n}, F., {et~al.} 2017, \rmxaa, 53,
  385

\bibitem[{{Ferland} {et~al.}(1998){Ferland}, {Korista}, {Verner}, {Ferguson},
  {Kingdon}, \& {Verner}}]{1998Ferland_pasp110}
{Ferland}, G.~J., {Korista}, K.~T., {Verner}, D.~A., {et~al.} 1998, \pasp, 110,
  761

\bibitem[{{Ferrarotti} \& {Gail}(2006)}]{2006Ferrarotti_aap447}
{Ferrarotti}, A.~S. \& {Gail}, H.-P. 2006, \aap, 447, 553

\bibitem[{{Fitzpatrick}(1999)}]{1999Fitzpatrick_pasp111}
{Fitzpatrick}, E.~L. 1999, \pasp, 111, 63

\bibitem[{{Forrest} {et~al.}(1981){Forrest}, {Houck}, \&
  {McCarthy}}]{1981Forrest_apj248}
{Forrest}, W.~J., {Houck}, J.~R., \& {McCarthy}, J.~F. 1981, \apj, 248, 195

\bibitem[{{Garc{\'{\i}}a-Hern{\'a}ndez}
  {et~al.}(2010){Garc{\'{\i}}a-Hern{\'a}ndez}, {Manchado},
  {Garc{\'{\i}}a-Lario}, {Stanghellini}, {Villaver}, {Shaw}, {Szczerba}, \&
  {Perea-Calder{\'o}n}}]{2010Garcia-Hernandez_apjl724}
{Garc{\'{\i}}a-Hern{\'a}ndez}, D.~A., {Manchado}, A., {Garc{\'{\i}}a-Lario},
  P., {et~al.} 2010, \apjl, 724, L39

\bibitem[{{Garc{\'{\i}}a-Hern{\'a}ndez}
  {et~al.}(2012){Garc{\'{\i}}a-Hern{\'a}ndez}, {Villaver},
  {Garc{\'{\i}}a-Lario}, {Acosta-Pulido}, {Manchado}, {Stanghellini}, {Shaw},
  \& {Cataldo}}]{2012Garcia-Hernandez_apj760}
{Garc{\'{\i}}a-Hern{\'a}ndez}, D.~A., {Villaver}, E., {Garc{\'{\i}}a-Lario},
  P., {et~al.} 2012, \apj, 760, 107

\bibitem[{{Goebel} \& {Moseley}(1985)}]{1985Goebel_apjl290}
{Goebel}, J.~H. \& {Moseley}, S.~H. 1985, \apjl, 290, L35

\bibitem[{{Grishko} {et~al.}(2001){Grishko}, {Tereszchuk}, {Duley}, \&
  {Bernath}}]{2001Grishko_apjl558}
{Grishko}, V.~I., {Tereszchuk}, K., {Duley}, W.~W., \& {Bernath}, P. 2001,
  \apjl, 558, L129

\bibitem[{{Hillier} \& {Miller}(1998)}]{1998Hillier_apj496}
{Hillier}, D.~J. \& {Miller}, D.~L. 1998, \apj, 496, 407

\bibitem[{{Hoare}(1990)}]{1990Hoare_mnra244}
{Hoare}, M.~G. 1990, \mnras, 244, 193

\bibitem[{{Hoare} {et~al.}(1992){Hoare}, {Roche}, \&
  {Clegg}}]{1992Hoare_mnra258}
{Hoare}, M.~G., {Roche}, P.~F., \& {Clegg}, R.~E.~S. 1992, \mnras, 258, 257

\bibitem[{{Hony} {et~al.}(2002){Hony}, {Waters}, \&
  {Tielens}}]{2002Hony_aap390}
{Hony}, S., {Waters}, L.~B.~F.~M., \& {Tielens}, A.~G.~G.~M. 2002, \aap, 390,
  533

\bibitem[{{Ishiki} {et~al.}(2018){Ishiki}, {Okamoto}, \&
  {Inoue}}]{2018MNRAS.474.1935I}
{Ishiki}, S., {Okamoto}, T., \& {Inoue}, A.~K. 2018, \mnras, 474, 1935

\bibitem[{{Jenkins}(2009)}]{2009Jenkins_apj700}
{Jenkins}, E.~B. 2009, \apj, 700, 1299

\bibitem[{{Kwok} \& {Zhang}(2011)}]{2011Kwok_nat}
{Kwok}, S. \& {Zhang}, Y. 2011, Nature

\bibitem[{{Lenzuni} {et~al.}(1989){Lenzuni}, {Natta}, \&
  {Panagia}}]{1989Lenzuni_apj345}
{Lenzuni}, P., {Natta}, A., \& {Panagia}, N. 1989, \apj, 345, 306

\bibitem[{{Liu} {et~al.}(2001){Liu}, {Barlow}, {Cohen}, {Danziger}, {Luo},
  {Baluteau}, {Cox}, {Emery}, {Lim}, \& {P{\'e}quignot}}]{2001Liu_mnra323}
{Liu}, X., {Barlow}, M.~J., {Cohen}, M., {et~al.} 2001, \mnras, 323, 343

\bibitem[{{Morisset}(2006)}]{2006Morisset_234}
{Morisset}, C. 2006, in IAU Symposium, Vol. 234, Planetary Nebulae in our
  Galaxy and Beyond, ed. {M.~J.~Barlow \& R.~H.~M{\'e}ndez}, 467--468

\bibitem[{{Morisset}(2014)}]{2014Morisset_}
{Morisset}, C. 2014, in Asymmetrical Planetary Nebulae VI conference,
  Proceedings of the conference held 4-8 November, 2013. Edited by C. Morisset,
  G. Delgado-Inglada and S. Torres-Peimbert.

\bibitem[{{Morisset} \& {Georgiev}(2009)}]{2009Morisset_aap507}
{Morisset}, C. \& {Georgiev}, L. 2009, \aap, 507, 1517

\bibitem[{{Moseley} \& {Silverberg}(1985)}]{1985NASCP2353..233M}
{Moseley}, H. \& {Silverberg}, R.~F. 1985, in NASA Conference Publication, Vol.
  2353, NASA Conference Publication, 233--239

\bibitem[{{Moseley} \& {Silverberg}(1986)}]{1986NASCP2403A..18M}
{Moseley}, H. \& {Silverberg}, R.~F. 1986, in NASA Conference Publication, Vol.
  2403, NASA Conference Publication, ed. J.~A. {Nuth}, III \& R.~E. {Stencel}

\bibitem[{{Murphy} {et~al.}(2010){Murphy}, {Sadler}, {Ekers}, {Massardi},
  {Hancock}, {Mahony}, {Ricci}, {Burke-Spolaor}, {Calabretta}, {Chhetri}, {de
  Zotti}, {Edwards}, {Ekers}, {Jackson}, {Kesteven}, {Lindley}, {Newton-McGee},
  {Phillips}, {Roberts}, {Sault}, {Staveley-Smith}, {Subrahmanyan}, {Walker},
  \& {Wilson}}]{2010Murphy402}
{Murphy}, T., {Sadler}, E.~M., {Ekers}, R.~D., {et~al.} 2010, \mnras, 402, 2403

\bibitem[{{Otsuka} {et~al.}(2014){Otsuka}, {Kemper}, {Cami}, {Peeters}, \&
  {Bernard-Salas}}]{2014Otsuka_mnra437}
{Otsuka}, M., {Kemper}, F., {Cami}, J., {Peeters}, E., \& {Bernard-Salas}, J.
  2014, \mnras, 437, 2577

\bibitem[{{Otsuka} {et~al.}(2017){Otsuka}, {Ueta}, {van Hoof}, {Sahai},
  {Aleman}, {Zijlstra}, {Chu}, {Villaver}, {Leal-Ferreira}, {Kastner},
  {Szczerba}, \& {Exter}}]{2017Otsuka_aaps231}
{Otsuka}, M., {Ueta}, T., {van Hoof}, P.~A.~M., {et~al.} 2017, \apjs, 231, 22

\bibitem[{Patnaik(2003)}]{patnaik2003handbook}
Patnaik, P. 2003, Handbook of Inorganic Chemicals, McGraw-Hill handbooks
  (McGraw-Hill)

\bibitem[{{Pegourie}(1988)}]{1988Pegourie_aap194}
{Pegourie}, B. 1988, \aap, 194, 335

\bibitem[{{Phillips} {et~al.}(1986){Phillips}, {Mampaso}, {Vilchez}, \&
  {Gomez}}]{1986Phillips_apss122}
{Phillips}, J.~P., {Mampaso}, A., {Vilchez}, J.~M., \& {Gomez}, P. 1986, \apss,
  122, 81

\bibitem[{{Planck Collaboration} {et~al.}(2015){Planck Collaboration},
  {Arnaud}, {Atrio-Barandela}, {Aumont}, {Baccigalupi}, {Banday}, {Barreiro},
  {Battaner}, {Benabed}, {Benoit-L{\'e}vy}, {Bernard}, {Bersanelli},
  {Bielewicz}, {Bonaldi}, {Bond}, {Borrill}, {Bouchet}, {Buemi}, {Burigana},
  {Cardoso}, {Casassus}, {Catalano}, {Cerrigone}, {Chamballu}, {Chiang},
  {Colombi}, {Colombo}, {Couchot}, {Crill}, {Curto}, {Cuttaia}, {Davies},
  {Davis}, {de Bernardis}, {de Rosa}, {de Zotti}, {Delabrouille}, {Dickinson},
  {Diego}, {Donzelli}, {Dor{\'e}}, {Dupac}, {En{\ss}lin}, {Eriksen}, {Finelli},
  {Frailis}, {Franceschi}, {Galeotta}, {Ganga}, {Giard}, {Gonz{\'a}lez-Nuevo},
  {G{\'o}rski}, {Gregorio}, {Gruppuso}, {Hansen}, {Harrison}, {Hildebrandt},
  {Hivon}, {Holmes}, {Hora}, {Hornstrup}, {Hovest}, {Huffenberger}, {Jaffe},
  {Jones}, {Juvela}, {Keih{\"a}nen}, {Keskitalo}, {Kisner}, {Knoche}, {Kunz},
  {Kurki-Suonio}, {L{\"a}hteenm{\"a}ki}, {Lamarre}, {Lasenby}, {Lawrence},
  {Leonardi}, {Leto}, {Lilje}, {Linden-V{\o}rnle}, {L{\'o}pez-Caniego},
  {Mac{\'{\i}}as-P{\'e}rez}, {Maffei}, {Maino}, {Mandolesi}, {Martin}, {Masi},
  {Massardi}, {Matarrese}, {Mazzotta}, {Mendes}, {Mennella}, {Migliaccio},
  {Miville-Desch{\^e}nes}, {Moneti}, {Montier}, {Morgante}, {Mortlock},
  {Munshi}, {Murphy}, {Naselsky}, {Nati}, {Natoli}, {Noviello}, {Novikov},
  {Novikov}, {Pagano}, {Pajot}, {Paladini}, {Paoletti}, {Peel}, {Perdereau},
  {Perrotta}, {Piacentini}, {Piat}, {Pietrobon}, {Plaszczynski},
  {Pointecouteau}, {Polenta}, {Popa}, {Pratt}, {Procopio}, {Prunet}, {Puget},
  {Rachen}, {Reinecke}, {Remazeilles}, {Ricciardi}, {Riller}, {Ristorcelli},
  {Rocha}, {Rosset}, {Roudier}, {Rubi{\~n}o-Mart{\'{\i}}n}, {Rusholme},
  {Sandri}, {Savini}, {Scott}, {Spencer}, {Stolyarov}, {Sutton}, {Suur-Uski},
  {Sygnet}, {Tauber}, {Terenzi}, {Toffolatti}, {Tomasi}, {Trigilio},
  {Tristram}, {Trombetti}, {Tucci}, {Umana}, {Valiviita}, {Van Tent}, {Vielva},
  {Villa}, {Wade}, {Wandelt}, {Zacchei}, {Zijlstra}, \&
  {Zonca}}]{2015Planck-Collaboration_aap573}
{Planck Collaboration}, {Arnaud}, M., {Atrio-Barandela}, F., {et~al.} 2015,
  \aap, 573, A6

\bibitem[{{Pottasch} {et~al.}(2004){Pottasch}, {Bernard-Salas}, {Beintema}, \&
  {Feibelman}}]{2004Pottasch_aap423}
{Pottasch}, S.~R., {Bernard-Salas}, J., {Beintema}, D.~A., \& {Feibelman},
  W.~A. 2004, \aap, 423, 593

\bibitem[{Ropp(2012)}]{ropp2012encyclopedia}
Ropp, R. 2012, Encyclopedia of the Alkaline Earth Compounds (Elsevier Science)

\bibitem[{{Rouleau} \& {Martin}(1991)}]{1991Rouleau_apj377}
{Rouleau}, F. \& {Martin}, P.~G. 1991, \apj, 377, 526

\bibitem[{{Shields}(1983)}]{1983Shields_103}
{Shields}, G.~A. 1983, in IAU Symposium, Vol. 103, Planetary Nebulae, ed. D.~R.
  {Flower}, 259--263

\bibitem[{{Sutherland} \& {Dopita}(2017)}]{2017Sutherland_apjs229}
{Sutherland}, R.~S. \& {Dopita}, M.~A. 2017, \apjs, 229, 34

\bibitem[{{Treffers} \& {Cohen}(1974)}]{1974Treffers_apj188}
{Treffers}, R. \& {Cohen}, M. 1974, \apj, 188, 545

\bibitem[{van~de Hulst(1957)}]{1957vandeHulst}
van~de Hulst, H.~C. 1957, Light scattering by small particles ({Chapman and
  Hall})

\bibitem[{{Volk} {et~al.}(1997){Volk}, {Dinerstein}, \&
  {Sneden}}]{1997Volk_180}
{Volk}, K., {Dinerstein}, H., \& {Sneden}, C. 1997, in IAU Symposium, Vol. 180,
  Planetary Nebulae, ed. H.~J. {Habing} \& H.~J.~G.~L.~M. {Lamers}, 284

\bibitem[{{Vollmer} {et~al.}(2010){Vollmer}, {Gassmann}, {Derri{\`e}re},
  {Boch}, {Louys}, {Bonnarel}, {Dubois}, {Genova}, \&
  {Ochsenbein}}]{2010Vollmer511}
{Vollmer}, B., {Gassmann}, B., {Derri{\`e}re}, S., {et~al.} 2010, \aap, 511,
  A53

\bibitem[{{Weingartner} {et~al.}(2006){Weingartner}, {Draine}, \&
  {Barr}}]{2006Weingartner_apj645}
{Weingartner}, J.~C., {Draine}, B.~T., \& {Barr}, D.~K. 2006, \apj, 645, 1188

\bibitem[{{Willner} {et~al.}(1979){Willner}, {Jones}, {Russell}, {Puetter}, \&
  {Soifer}}]{1979Willner_apj234}
{Willner}, S.~P., {Jones}, B., {Russell}, R.~W., {Puetter}, R.~C., \& {Soifer},
  B.~T. 1979, \apj, 234, 496

\bibitem[{{Wright} {et~al.}(1991){Wright}, {Wark}, {Troup}, {Otrupcek},
  {Jennings}, {Hunt}, \& {Cooke}}]{1991Wright251}
{Wright}, A.~E., {Wark}, R.~M., {Troup}, E., {et~al.} 1991, \mnras, 251, 330

\bibitem[{{Zhang} {et~al.}(2009){Zhang}, {Jiang}, \& {Li}}]{2009Zhang_apj702}
{Zhang}, K., {Jiang}, B.~W., \& {Li}, A. 2009, \apj, 702, 680

\bibitem[{{Zhukovska} \& {Gail}(2008)}]{2008Zhukovska_aap486}
{Zhukovska}, S. \& {Gail}, H.-P. 2008, \aap, 486, 229

\end{thebibliography}

\begin{appendix}

\section{Degeneracy between dust size and distance to the heating source}\label{append:degeneracy}

\begin{figure}
\centering
\includegraphics[width=9cm]{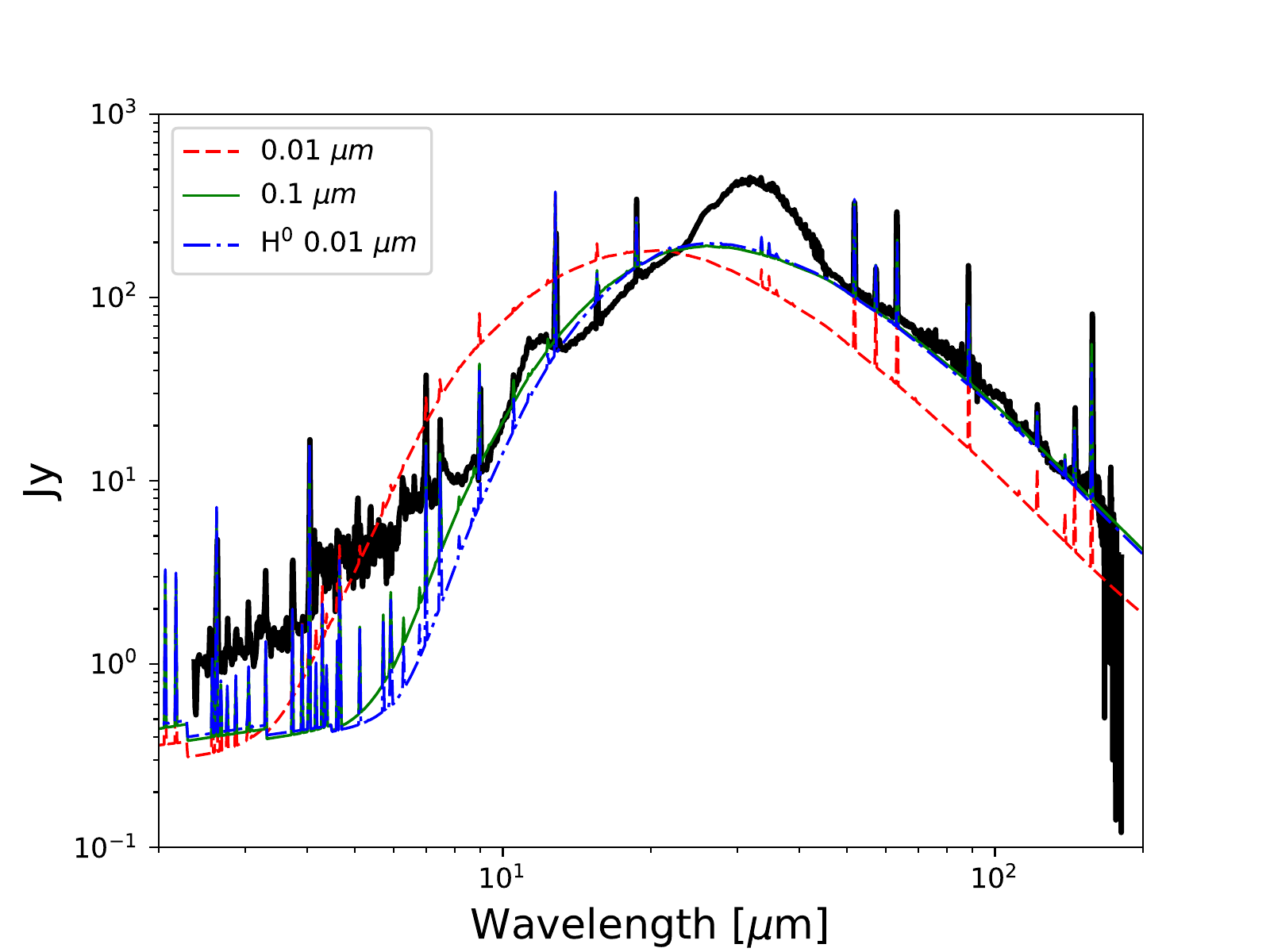}
\caption{Observed ISO spectrum (Black) and results of models with small dust (Red dashed), bigger grains (Green solid) and small grains only in the PDR (Blue dash-doted).}
\label{fig:degen}%
\end{figure}

To explore the role of the grain size and the distance of the grain to the central source, we run some toy models using single size BE1 amorphous carbon dust grains. A model with canonical size of 0.01~\mm is shown in red in Fig.~\ref{fig:degen}. The emission is clearly too hot (peaking at to short wavelengths). 

To reduce the temperature of the grains and shift the emission to longer wavelengths, we tested two options: 1) increasing the size of the grains (0.1~\mm, green line) and 2) having the dust the canonical size, but distributed only in the PDR (blue line in Fig.~\ref{fig:degen}). This latest option is achieved by applying a scale factor to the dust abundance. In our case, the scale factor is the hydrogen atomic fraction, small when gas is ionized and unity when atomic. 
The two options clearly move the emission toward the longer wavelengths, allowing the fit between 50 and 200~\mm to be quite good and leading to a degeneracy between the size of the grains and their position in the nebula. 

An interesting result of exploring this degeneracy is that in both cases, the total dust mass required to obtain two very close emissions are very similar.
It is difficult to really prefer one of the two options only comparing the fits, and the real situation may also be any smooth combination of both of them: grains being bigger and denser (in terms of dust-to-gas ratio) with the increasing distance to the star. For the final model, we choose to combine both cases: big grains, but only in the H$^0$ region.

\end{appendix}

\end{document}